\documentstyle[12pt,aaspp4]{article}

\def\linebreak{\hfil\break}

\def\
{\hfil\linebreak}
\def\singlespace{\baselineskip=14pt}

\def\singlespace{%
    \lineskip                .15ex
    \baselineskip            3.0ex
   \lineskiplimit              0ex
   \parskip                0.60ex plus .30ex minus .15ex
   }%

\def\etal{{\it et al}. }
\def\degree{\ifmmode {^\circ}\else {$^\circ$}\fi}
\def\mum{\ifmmode {\rm \mu {\rm m}}\else $\rm \mu {\rm m}$\fi}
\def\arcsec{\ifmmode ^{\prime \prime}\else $^{\prime \prime}$\fi}

\def\inch{\ifmmode ^{\prime \prime}\else $^{\prime \prime}$\fi}
\def\arcmin{\ifmmode ^{\prime}\else $^{\prime}$\fi}

\def\msun{\ifmmode {\rm M_{\odot}}\else $\rm M_{\odot}$\fi}
\newbox\grsign \setbox\grsign=\hbox{$>$} \newdimen\grdimen \grdimen=\ht\grsign
\newbox\simlessbox \newbox\simgreatbox
\setbox\simgreatbox=\hbox{\raise.5ex\hbox{$>$}\llap
     {\lower.5ex\hbox{$\sim$}}}\ht1=\grdimen\dp1=0pt
\setbox\simlessbox=\hbox{\raise.5ex\hbox{$<$}\llap
     {\lower.5ex\hbox{$\sim$}}}\ht2=\grdimen\dp2=0pt

\begin{document}

\singlespace

\centerline{\Large {\bf Gravitational Stirring in Planetary Debris Disks}}
\vskip 7ex
\centerline{Scott J. Kenyon}
\centerline{Smithsonian Astrophysical Observatory}
\centerline{60 Garden Street, Cambridge, MA 02138} 
\centerline{e-mail: skenyon@cfa.harvard.edu}
\vskip 3ex
\centerline{and}
\vskip 3ex
\centerline{Benjamin C. Bromley}
\centerline{Department of Physics}
\centerline{University of Utah}
\centerline{201 JFB, Salt Lake City, UT 84112} 
\centerline{e-mail:bromley@physics.utah.edu}
\vskip 7ex
%
\centerline{to appear in}
\centerline{{\it The Astronomical Journal}}
\centerline{January 2001}
\vskip 5ex
\received{27 June 2000}
\accepted{12 September 2000}

\clearpage

%
%
\singlespace

\begin{abstract}

We describe gravitational stirring models of planetary debris disks
using a new multi-annulus planetesimal evolution code.  The current
code includes gravitational stirring and dynamical friction; future
studies will include coagulation, fragmentation, Poynting-Robertson
drag, and other physical processes.  We use the results of our
calculations to investigate the physical conditions required for
small bodies in a planetesimal disk to reach the shattering velocity
and begin a collisional cascade.  Our results demonstrate that disks 
composed primarily of bodies with a single size will not undergo a 
collisional cascade which produces small dust grains at 30--150 AU
on timescales of 1 Gyr or smaller.  Disks with a size distribution
of bodies reach conditions necessary for a collisional cascade in
10 Myr to 1 Gyr if the disk is at least as massive as a minimum mass 
solar nebula and if the disk contains objects with radii of 500 km 
or larger.  The estimated $\sim$ 500 Myr survival time for these 
disks is close to the median age of $\sim$ 400 Myr derived for 
nearby stars with dusty disks.

\end{abstract}

\subjectheadings{planetary systems -- solar system: formation --
stars: formation -- circumstellar matter}

\section{INTRODUCTION}

Planetary debris disks surround nearly all stars during 
the early stages of their main sequence lifetimes.
Recent surveys with the {\it Infrared Space Observatory (ISO)}
indicate a median disk lifetime of $\sim$ 400 $\times ~ 10^6$ yr 
(Myr hereafter; \cite{bec98}; \cite{dom98}; \cite{gai99}; 
\cite{hab99}; \cite{faj99}; \cite{son00}).  
The disk fraction decreases dramatically with age and is much less 
than 10\% for stars with ages of $10^9$ yr (Gyr hereafter) or larger.  

Because thermal emission from cold dust is large compared to the 
photospheric radiation from the central star, most debris disks 
have been detected with mid-infrared or far-infrared observations
(\cite{bac93}; \cite{art97}; \cite{lag00}).  
Infrared and optical images reveal a disk-like morphology with 
an outer radius ranging from 100--200 AU up to 1000 AU in several 
systems (\cite{au99a}, 1999b; \cite{gre98}; \cite{hol98};
\cite{jay98}, 2000; \cite{koe98}; \cite{sch99}; \cite{tri98}; \cite{tri00}).
The disk often has a central `hole,' where radiation from small
dust grains is less than emission from material in the outer disk.
The ratio of disk to stellar luminosity is 
$L_d/L_{\star} \sim 10^{-5}$ to $10^{-2}$ (\cite{art97}; 
Fajardo-Acosta, Stencel, \& Backman 1997; \cite{faj99}; \cite{lag00}).
The disks have a small gas content, with measured gas-to-dust
mass ratios of $M_g/M_d \lesssim$ 1--10 (\cite{zuc95}; \cite{den95};
Lecavelier des Etangs, Vidal-Madjar, \& Ferlet 1998;
Greaves, Coulson, \& Holland 2000).  
Models for the scattered and thermal emission suggest grains with 
sizes of 1--100 $\mu$m and a total mass in small grains of $\sim$ 
0.01 $M_E$, where 1 $M_E = 6 \times 10^{27}$ g is the mass of the 
earth (e.g., Greaves, Mannings, \& Holland 2000). 

Theoretical models for a debris disk begin with dust grains 
in Keplerian orbits about the central star (see \cite{art97}, 
\cite{lag00}, and references therein).
For particle sizes of 1--100 \mum, Poynting-Robertson drag and 
radiation pressure remove dust from the disk in $\sim $ 1--10 Myr 
(\cite{bur79}; \cite{bac93}; Backman, Dasgupta, \& Stencel 1995; 
\cite{art97}). 
Collisions between larger bodies can replenish small grains if 
the collision velocity is $\sim$ 100--300 m s$^{-1}$ (see below). 
These large velocities
initiate a ``collisional cascade,'' where planetesimals with radii
of 1--10 km are ground down into smaller and smaller bodies.
A collisional cascade requires a mass reservoir of $\sim$ 10--100 $M_E$
to replenish smaller grains over a disk lifetime of 100 Myr or more 
(\cite{art97}; \cite{hab99}).  
This mass is similar to estimates for the original planetesimal mass 
in the Kuiper Belt of our Solar System (e.g., \cite{ste95}, 1996;
Kenyon \& Luu 1998 [KL98], 1999 [KL99]).  

Despite the popularity of this picture, the origin of the large 
collision velocities is uncertain.  Because the difference between 
the gas velocity and dust velocity is small and circularization 
is efficient, dust grains and larger bodies within a protosolar 
nebula probably had nearly circular orbits initially (see 
KL99 and references therein).  
Short-term encounters with passing stars and stirring by planets 
embedded in the disk can increase the velocities of dust grains (e.g., 
\cite{art89}; \cite{mou97}; \cite{ida00}; \cite{kal00}).
Although stellar encounters can increase particle velocities
enormously, such encounters are probably rare.  
Collisions of bodies within the disk may also effectively damp 
large velocities following the encounter; KL99 estimate damping 
times of 1--10 Myr for 1--100 m objects during the early stages 
of planetesimal growth in the Kuiper Belt. 
Stirring by embedded planets is attractive, because objects with
radii of 1000 km or more naturally form in the disk and these
objects continuously stir up the velocities of small dust grains.

The timescale to reach the large collision velocities required 
in many debris disks is uncertain.  Kenyon \etal (1999) show 
that Pluto-sized objects can form within the dusty ring of HR4796 A 
and stir up the velocities of smaller objects in $\sim$ 10 Myr 
if the disk is sufficiently massive.  Gravitational stirring may 
be less efficient in lower mass disks (KL99) and in the outer 
portions of the debris disks of $\beta$ Pic and other systems.

Our goal is to investigate the physical conditions needed in a
planetary debris disk for embedded planets to stir up smaller 
objects to the shattering velocity.  We use a new multi-annulus 
planetesimal evolution code to compute the dynamical evolution 
of small bodies and show that results for simple test problems 
agree with previous calculations.  Our 
calculations for debris disks with masses comparable to the
`minimum mass solar nebula' indicate that small bodies with radii 
of 200 km or less can stir much smaller bodies in the disk on 
timescales of 1 Gyr or longer.  Shorter stirring timescales 
require larger objects and more massive disks.  Objects with radii of 
500 km or larger can stir low mass planetesimals to the shattering
limit in disks with 5--10 times the mass of a minimum mass solar
nebula in 10--20 Myr, close to the inferred ages for $\beta$ Pic 
and HR 4796A (\cite{bar97}, 1999; see also \cite{son00}).  
Once small bodies reach the shattering limit, the lifetime of
small bodies at 100 AU is roughly 500 Myr for a minimum mass solar
nebula.  These results suggest that planetesimal growth in protosolar 
nebulae with a range of initial masses can account for the observed 
range of ages of planetary debris disk systems.

We outline the model in \S2, describe the calculations in \S3,
and conclude with a brief discussion in \S4. 

\section{The Model}

As in KL98 and KL99, we adopt Safronov's (1969) particle-in-a-box method, 
where we treat planetesimals as a statistical ensemble of masses with a 
distribution of horizontal and vertical velocities about a Keplerian 
orbit\footnote{Kokubo \& Ida (1996) compare several results from this 
statistical theory to results of direct $n$-body calculations.}.  
The model grid contains $N$ concentric annuli centered at heliocentric 
distances $a_i$.  Each annulus has an inner radius at 
$a_i - \delta a_i/2$ and an outer radius at $a_i + \delta a_i/2$.  
The midpoint of the model grid is at a heliocentric distance $a_{mid}$.  
Calculations begin with a differential mass distribution $n(m_i$) 
of bodies with horizontal and vertical velocity dispersions $h_i(t)$ 
and $v_i(t)$.  We approximate the continuous distribution of particle 
masses with discrete batches having particle populations $n_i(t)$ and 
total masses $M_i(t)$.  

To evolve the velocity distribution in time, we solve the Fokker-Planck
equation for an ensemble of masses undergoing dynamical friction and
viscous stirring.  Dynamical friction transfers kinetic energy from
large bodies to small bodies and drives a system to energy equipartition.
Viscous stirring converts energy from the stellar potential into
random motion and increases the velocities of all planetesimals.
We do not allow physical collisions in this 
study; all gravitational interactions are elastic and do not
modify the initial mass distribution.  The Appendices of KL98 and 
KL99 describe accretion and velocity evolution for planetesimals 
in a single annulus and compare numerical results with analytic 
solutions for standard test cases.  The Appendix of this paper 
generalizes KL98 and KL99 for a disk with $N$ concentric annuli 
and describes several test cases (see also \cite{spa91}; \cite{wei97}).

The starting conditions for these calculations are based on KL99 
and Kenyon \etal (1999).  We consider systems of $N$ annuli in disks
with $a_{mid}$ = 35--140 AU and $\delta a_i/a_i$ = 0.001--0.0125. 
The central star has a mass of 3 \msun, appropriate for debris 
disks surrounding nearby A-type stars.  Timescales for other central 
stars and other heliocentric distances scale with stellar mass and
heliocentric distance as $(a^3/M_{\star})^{1/2}$. 
The particles begin with eccentricity $e_0$ and inclination $i_0 = e_0/2$.
The initial velocities are then $h_{i0} = 0.79 ~ e_0 ~ V_{Ki}$ and 
$v_{i0} = 0.71 ~ {\rm sin} ~ i_0 ~ V_{Ki}$ where $V_{Ki}$ is the 
velocity of a circular orbit centered on annulus $i$. Several tests 
show that the timescale to reach large random velocities is fairly 
independent of $e_0$ for $e_0 \lesssim 10^{-3}$.

\section{Calculations}

\subsection{A Simple Model for the Disk Lifetime}

To assess the outcome of the stirring calculations, we identify 
models which can initiate a collisional cascade and produce enough
debris to account for observations of currently known debris disk
systems.  The debris produced in a collision between two bodies 
of mass $m_i$ and $m_j$ is set by the center-of-mass collision energy,

\begin{equation}
Q_{c,ij} = \frac{0.25~m_i~m_j~V_{I,ij}^2}{(m_i + m_j)^2}
\end{equation}

\noindent
where $V_{I,ij}$ is the impact velocity.
We adopt $V_{I,ij}^2$ = $V_{ij}^2 + V_{e,ij}^2$,
where $V_{ij}$ is the relative velocity of the colliding planetesimals
and $V_{e,ij}^2$ = $2 G (m_i + m_j)/(r_i + r_j)$ is the mutual escape
velocity (see the Appendix of KL99 and references therein).
The energy needed to fragment the colliding bodies and eject a fraction, 
$f_{ej}$, of their combined mass is 

\begin{equation}
Q_f = S ~ \left ( \frac{\alpha_V~f_{ej}^{1~+~2/\alpha_V}}{f_{KE} (\alpha_V - 2)} \right ) ~ .
\end{equation}

\noindent
where $S$ is the binding energy of the combined mass\footnote{The binding
energy, $S$, is composed of an intrinsic tensile strength, $S_0$, and
a gravitational binding energy.  For objects with radii of 10 km or less,
$S \approx S_0$; $S$ is comparable to the gravitational binding energy 
for objects with radii of 100 km or larger.} (\cite{dav85}; 
Davis, Ryan, \& Farinella 1994).  
This expression assumes the ejected fragments receive a fixed fraction of 
the impact energy, $f_{KE} Q_{f,ij}$, and have a power law velocity 
distribution,

\begin{equation}
f(>v) \propto (v/v_c)^{-\alpha_V} ~ .
\end{equation}

\noindent
If $Q_f = Q_{c,ij}$ and $\alpha_V = 2.25$ (e.g., Davis \etal 1985; KL99), 
the impact velocity needed to eject $f_{ej} (m_i + m_j)$ is 

\begin{equation}
V_{I,ij} = 6 \left ( \frac{f_{ej}^{17/9} S}{f_{KE}} \right )^{1/2} ~ \left ( \frac{m_i + m_j}{\sqrt{m_im_j}} \right ) ~ .
\end{equation}

\noindent
This expression reduces to

\begin{equation}
V_{I,ij} \approx 800 ~ {\rm m ~ s^{-1}} ~ f_{ej}^{17/18} ~ ,
\end{equation}

\noindent
for a collision between two equal mass bodies with $S$ = 
$2 \times 10^6$ erg g$^{-1}$ and $f_{KE}$ = 0.05, typical values 
for small icy objects in the outer disk (see KL99 and references
therein).  For 1 km bodies with 
$m = 6.3 \times 10^{15}$ g, an impact velocity of 90 m s$^{-1}$ 
ejects $\sim$ 10\% of the combined mass.  An impact velocity of 
400 m s$^{-1}$ ejects $\sim$ 50\% of the combined mass.

To estimate the lifetime of a disk of mass $M_d$ undergoing a 
collisional cascade, we use the coagulation equation (see KL98
and references therein):

\begin{equation}
\frac{dn_i}{dt} \approx \left ( \frac{n_i n_j}{4 H a \delta a} \right ) ~V_{ij}~(r_i + r_j)^2
\end{equation}

\noindent
where $r_i$ and $r_j$ are the radii of the two bodies, 
$V_{ij}^2 = V_i^2 + V_j^2$ is the relative velocity, 
$H$ is the vertical scale height, and $\delta a$ is the width
of an annulus centered at $a$.
This expression neglects gravitational focusing.
The production rate of debris is $\dot{m} \equiv dm/dt = f_{ej} (m_i + m_j) dn/dt$.
The disk lifetime is then $\tau_d = M_d / \dot{m}$.  We adopt a disk
model with a surface density $\Sigma$ = $\Sigma_0 (a/a_0)^{-3/2}$
where $a_0$ = 1 AU. The disk mass is $m_d = 2 \pi \Sigma a \delta a$.
Collisions between equal mass bodies yields $n_i n_j = m_d^2/2m_i^2$;
the disk lifetime is then

\begin{equation}
\tau_d = \frac{H r_i \rho}{3 \Sigma_0 f_{ej} V_{ij}} \left ( \frac{a}{a_0} \right ) ^{3/2}
\end{equation}

\noindent
Substituting H = 0.64 $V_{ij} / \Omega_K$, where $\Omega_K$ is
the orbital frequency yields

\begin{equation}
\tau_d = 500 ~ {\rm Myr} ~ \left ( \frac{60 ~ {\rm g ~ cm^{-2}}}{\Sigma_0} \right ) \left ( \frac{r_i}{1 ~ {\rm km}} \right ) \left ( \frac{\rho}{1.5 ~ {\rm g ~ cm^{-3}}} \right ) \left ( \frac{0.2}{f_{ej}} \right ) \left ( \frac{a}{100 ~ {\rm AU}} \right )^3
\end{equation}

\noindent
where $\rho$ is the mass density of a planetesimal and
$\Sigma_0$ = 60 ${\rm g ~ cm^{-2}}$ is the surface density 
of the `minimum mass solar nebula' (\cite{hay81}; \cite{wei77}).
We adopt $r_i$ = 1 km, which is roughly the peak of the mass function for
coagulation calculations in the outer disk (KL99; \cite{ken99}).
The observed disk lifetime of 300--500 Myr for most systems implies
$f_{ej} \approx$ 0.2 and an impact velocity of $\sim$ 175 m s$^{-1}$.
The relative velocities of the 1 km planetesimals are then $\sim$ 125 m s$^{-1}$.

A disk with $V_{ij}$ = 100--200 m s$^{-1}$ and $f_{ej}$ = 0.2
satisfies current constraints on the thermal emission and
scattered light component of debris disks.  For a disk with scale
height $H$, the solid angle of the disk as seen from the central 
star is $\Omega/4 \pi$ = $H/2a$ $\approx$ 0.01--0.02 for $a$ = 100 AU
and $V_{ij}$ = 100--200 m s$^{-1}$.  The thermal luminosity from
the disk $L_{th}$ and the scattered luminosity from the disk
$L_{sc}$ depend on the radial optical depth $\tau$ and the
grain albedo $\omega$:
$L_{th}/L_{\star} = \tau (1 - \omega) (H/2a)$
and
$L_{sc}/L_{\star} = \tau \omega (H / 2 a).$
The observed range of luminosity ratios,
$L_{th}/L_{\star}$ $\approx$ $L_{sc}/L_{\star}$ $\approx$
$10^{-5}$ to $10^{-2}$ requires 
$\tau \approx$ $10^{-3}$ to 1 for
$\omega$ = 0.1--0.3 (see also
\cite{art89}; \cite{bac93}; \cite{ken99})

These estimates suggest that collisions of 1 km bodies in a 
circumstellar disk will yield an observable amount of debris 
over a disk lifetime of 100 Myr to 1 Gyr.  The 1 km planetesimals
must have relative velocities of 100 m s$^{-1}$ or more and
reside in a disk with an initial mass close to or larger than
the minimum mass solar nebula model.  We now consider whether 
gravitational stirring in such a disk allows 1 km bodies to 
reach such large impact velocities.

\subsection{Disks with Single Mass Planetesimals}

To understand how gravitational stirring modifies particle velocities 
in a debris disk, we first consider a small portion of a disk surrounding
a 3 \msun~star.  The disk is composed of bodies with a single mass $m_0$ and 
a surface density $\Sigma(a)$ = 0.1 g cm$^{-2}$ $x ~ (a/{\rm 70~AU})^{-3/2}$,
where $x$ is a dimensionless constant.  A minimum mass solar nebula
has $x$ = 1 and $\Sigma$(70 AU) $\approx$ 0.1 g cm$^{-2}$. 
We restrict our study to bodies with r = 1--100 km; 
$m_0 = 6.3 \times 10^{15}$ g
to $6.3 \times 10^{21}$ g for icy objects with a mean density of
1.5 g cm$^{-3}$.  Smaller bodies have negligible stirring on cosmological
timescales; larger objects are nearly impossible to shatter (KL98; KL99).
Each body has $e_0 = 10^{-3}$ and $i_0 = 5 \times 10^{-4}$.
We calculate the velocity evolution for $x$ = $10^{-2}, 10^{-1}$, 1, and 
$10^1$ in model grids composed of 16 annuli with $\delta a / a$ = 0.001
and centered at $a_{mid}$ = 35, 70, and 140 AU.  This study illustrates 
how viscous stirring modifies planetesimal
velocities; dynamical friction occurs only between bodies in adjacent
annuli and is small.

Figures 1 and 2 show the evolution of the average particle velocity,
$V_i = (h_i^2 + v_i^2)^{1/2}$, and the vertical scale height,
$H_i = \sqrt{2} v_i / \Omega_{Ki}$, for the central four annuli of the 
model grid.  A collection of 1 km objects does not evolve in
1 Gyr at 35--140 AU.  More massive bodies evolve on an observable 
timescale.  The velocity and scale height for 10 km bodies increase 
modestly after 1 Gyr, but do not reach the physical conditions 
needed to begin a collisional cascade (Figure 1).  The velocity at
1 Gyr is 
$V_i$(1 Gyr) $\approx$ 13 m s$^{-1}$ $x^{1/4}$ $(a/{\rm 70~AU})^{-3/4}$.
The scale height of these objects also reaches a small fraction, 
$\sim$ 10\%, of the value needed to reprocess a significant fraction 
of the stellar luminosity.  Calculations with 100 km bodies are 
more encouraging.  In a massive disk, these larger bodies stir themselves 
rapidly and reach 100 m s$^{-1}$ velocities on 100 Myr timescales.  
The velocity at 1 Gyr again depends on $a$ and $x$, $v$(1 Gyr) 
$\approx$ 60 m s$^{-1}$ $x^{1/4}$ $(a/{\rm 70~AU})^{-3/4}$.  Thus, 
in disks with masses at least as large as a minimum mass solar nebula,
it is possible to reach velocities of 100 m s$^{-1}$ with 
gravitational stirring of 100 km bodies.

To test this conclusion further, we consider complete disk models
composed of single mass objects.  The disk consists of 128 annuli with
$\Delta a_i/a_i$ = 0.0125, and extends from $a_{in}$ = 30 AU
to $a_{out}$ = 150 AU. The surface density of planetesimals is
$\Sigma(a) = 0.1 (a/{\rm 70~AU})^{-3/2}$ g cm$^{-2}$, with
$e_0 = 10^{-3}$ and $i_0 = 5 \times 10^{-4}$.

Figures 3--4 show the evolution of the velocity and scale height in
each disk.  A disk with 10 km planetesimals evolves slowly during 1 Gyr.
The velocities reach 15\% of the shattering velocity; the
vertical scale height is sufficient to intercept only $\sim$ 0.02\%
of the stellar luminosity (Figure 3).  A disk with 100 km planetesimals 
achieves 100 m s$^{-1}$ velocities over 30--45 AU in 1 Gyr (Figure 4). 
The scale height is then large enough to intercept 0.1--0.2\% of the 
stellar luminosity.  As indicated by the symbols in each panel of Figure 3, 
the model with 10 km planetesimals yields velocities only $\sim$ 
10\% larger than the narrow grid model of Figure 1.  This difference 
grows to $\sim$ 20\% for 100 km planetesimals at 1 Gyr (Figure 4).  
Long-range stirring produces this velocity difference; large bodies exert 
more long-range gravitational forces on their distant neighbors 
than small bodies.  Both models 
produce a power law velocity distribution, $V_i \propto a^{-3/4}$, 
except at the edges of the grid.  The scale height increases slowly 
with radius, $H_i \propto a^{3/4}$.

Despite fairly rapid stirring of 100 km bodies, it is difficult for
these objects to initiate a collisional cascade.  The gravitational
binding energy of a 100 km object increases the shattering velocity 
by a factor of 2--3 (equation (4)).  Figures 3--4 show that velocities
of 200 m s$^{-1}$ cannot be reached in 1 Gyr or less.  Thus, a disk
composed primarily of planetesimals of a single size cannot undergo a
collisional cascade on a timescale of 1 Gyr or smaller.  

An obvious and more realistic alternative to this model is a disk 
composed of objects with a large range of sizes.  A set of 100 km 
bodies can effectively stir up smaller objects on timescales comparable 
to self-stirring models. These smaller objects can then initiate 
the collisional cascade.  This process would preserve the stirring 
population, because the large objects are harder to shatter than 
small objects.  We now consider such models to derive the timescale 
to reach the collisional cascade for a disk with a size distribution 
of planetesimals.

\subsection{Disks with a Size Distribution of Planetesimals}

As in the previous section, we begin with the velocity evolution of 
particles in small portions of a disk containing a size distribution 
of bodies with radii $r_i$ = 1 m to 500 km and surface density 
$\Sigma(a)$ = 0.1 g cm$^{-2}$ $x ~ (a/{\rm 70~AU})^{-3/2}$. We assume 
equal mass in each of 6 mass batches. This mass distribution is a 
natural outcome of coagulation calculations (KL99; Kenyon \etal 1999). 
The mass spacing factor between successive 
batches is $\delta = m_{i+1}/m_i$ = $10^3$.  Each body has $e_0 = 10^{-3}$ 
and $i_0 = 5 \times 10^{-4}$.  We calculate the velocity evolution 
for $x$ = 1 and $x$ = 10 in grids of 16 annuli with $\Delta a_i / a_i$
= 0.001 centered at $a_{mid}$ = 
35, 70, and 140 AU.  This study illustrates the timescale for 
dynamical friction to couple with viscous stirring to produce a 
swarm of small objects with eccentric orbits.

Figure 5 shows the evolution of $V_i$ for small objects in the 
central four annuli of the model grid.  Objects with $r$ = 1--1000 m
have nearly identical velocities; larger objects have velocities a factor 
of two or more smaller.  We label each curve with the radius (in km) 
of the largest object in the mass distribution, $r_{max}$.  The left panels 
show the velocity evolution for $x$ = 1; the right panels show the velocity 
evolution for $x$ = 10.  The velocities at 100 Myr to 1 Gyr are

\begin{equation}
V_i ({\rm 100 ~ Myr ~ to ~ 1 ~ Gyr}) \approx 25-41 \left ( \frac{r}{{\rm 100 ~ km}} \right )^{0.65} ~ \left ( \frac{a}{{\rm 70 ~ AU}} \right )^{-0.70}
\end{equation}

\noindent
for $x$ = 1 and

\begin{equation}
V_i ({\rm 100 ~ Myr ~ to ~ 1 ~ Gyr}) \approx 40-65 \left ( \frac{r}{{\rm 100 ~ km}} \right )^{0.65} ~ \left ( \frac{a}{{\rm 70 ~ AU}} \right )^{-0.70}
\end{equation}

\noindent
for $x$ = 10.  These equations fit the model velocities to 5\% or better.
Planetesimals with $r \gtrsim$ 500 km stir nebular material to the 
shattering limit in 1 Gyr or less.  Smaller planetesimals with
$r \sim$ 200 km can stir up the inner parts of a massive solar nebula,
but cannot stir the outer portions; these planetesimals also fail to 
stir a minimum mass solar nebula in 1 Gyr or less.  

We next consider complete disk models composed of a size distribution
of planetesimals. As in \S3.1, the disk has 128 annuli with $\Delta a_i/a_i$ 
= 0.0125, and extends from $a_{in}$ = 30 AU to $a_{out}$ = 150 AU. 
The surface density of planetesimals is $\Sigma(a) = 
0.1 (a/{\rm 70~AU})^{-3/2}$ g cm$^{-2}$, with
$e_0 = 10^{-3}$ and $i_0 = 5 \times 10^{-4}$.
Disk material is distributed among six mass batches with equal mass
per bin and $\delta = 10^3$.

Figures 6--7 show the evolution of the velocity and scale height in
disks with $r_{max}$ = 100 km (Figure 6) and 500 km (Figure 7).  
Calculations with $r_{max}$ = 100 km produce small objects with
velocities at the shattering limit in 1 Gyr.  These small bodies
are confined to a narrow range of disk radii, $a \lesssim$ 35--40 AU;
bodies outside these annuli have much smaller velocities.  Calculations
with $r_{max}$ = 500 km produce small objects with velocities at the 
shattering limit in roughly 10 Myr at 35 AU, in roughly 100 Myr at 70 AU,
and in less than 1 Gyr at 140 AU.  The disk with 500 km objects 
intercepts $\sim$ 0.1\% of the stellar flux at 50--100 Myr; disks with
100 km objects are detectable only after 0.5--1.0 Gyr. As indicated by 
the symbols in each panel, both of these models produce larger velocity 
objects compared to the narrow grid models of Figure 5. Long-range 
stirring produces this velocity difference. Both models also produce 
a power law velocity distribution, $V_i \propto a^{-3/4}$, except at 
the edges of the grid.  The scale height increases slowly with radius, 
$H_i \propto a^{3/4}$, as in the models with single-size planetesimals 
described above.

To test the sensitivity of the derived timescales to the mass distribution,
we calculated model grids with finer mass resolution. Figure 8 shows the 
velocity evolution for mass distributions at 35 AU with $x$ = 10 and 
$r_{max}$ = 500 km.  We label each curve in the figure with the mass 
spacing factor $\delta$.  Models with better mass resolution take longer 
to reach the shattering limit.  Because each model has the same $dN/dm$,
models with finer mass resolution have fewer of the largest bodies
than models with coarse mass resolution.  Thus, our computational
procedure produces smaller stirring rates in more finely spaced grids.
The difference in the velocity of small particles at 100 Myr as a 
function of mass spacing is modest: 
$V_i$ decreases from $\sim$ 190 m s$^{-1}$ for $\delta$ = $10^3$ to
$\sim$ 144 m s$^{-1}$ for $\delta$ = 5.6 and 
$\sim$ 132 m s$^{-1}$ for $\delta$ = 2.

\subsection{Limitations of the Models}

In this paper, we have begun to address some of the limitations and 
uncertainties of statistical calculations of planetesimal growth in 
a solar nebula (see KL98 and KL99 for a summary of these limitations).  
Our multi-annulus code should provide a better 
treatment of velocity evolution within the nebula, because we 
calculate short-range and long-range stirring using an evolving
density of planetesimals at each radius within the grid.  
Uncertainties due to the finite spatial resolution and the finite 
extent of the full grid are small (Figures 3--4, 6--7, and 9).  
These errors grow when we consider a small portion of a disk, 
as shown by the symbols in Figures 6--7.  Partial disk models with
large bodies, $r_{max} \gtrsim$ 200 km, have larger uncertainties 
than models with small bodies, $r_{max} \lesssim$ 200 km.  Particle
velocities are also sensitive to the mass resolution of the grid 
(Figure 8), but the errors are $\lesssim$ 50\% for any mass 
resolution, $\delta \lesssim 10^3$.  Larger errors are possible 
when the calculations include mergers and fragmentation (see 
KL98, KL99, and references therein).  We plan to examine this
issue for the multi-annulus evolution code in a future paper.

Our conclusions concerning the long term evolution of particle 
velocities are probably insensitive to our neglect of important 
physical processes, such as mergers with fragmentation and
radiative processes such as Poynting-Robertson drag.  The 
timescale to produce 100--1000 km objects in a solar nebula 
is generally short compared to the stirring time (KL99;
\cite{ken99}). Our assumption of an initial population of 
100--500 km objects therefore does not lead to a large 
uncertainty in the stirring timescale.  Previous calculations 
have also shown that collisions tend to redistribute kinetic 
energy from the larger bodies to the smaller bodies (KL99).  
Including this process in the calculations would probably
reduce stirring times by 25\% to 50\%.  Calculations with the
better mass resolution needed to follow mergers and fragmentation
tend to increase stirring times by a similar factor (Figure 8).
Thus, a more accurate treatment with better mass resolution 
which includes mergers and fragmentation will probably yield
stirring times close to those derived here.

Our neglect of processes which remove particles from the grid 
probably also has little impact on the derived stirring timescales.
Gas drag removes few of the 1 km objects needed for the collisional
cascade and has negligible impact on their velocities (KL99).
Poynting-Robertson drag and ejection by radiation pressure are
important for small particles once the collisional cascade begins
(\cite{bur79}; Backman, Dasgupta, \& Stencel 1995).  We have
concentrated on the evolution leading up to the collisional
cascade and plan to consider the cascade itself in a future paper.

Finally, our stirring timescales are upper limits for debris disks
containing larger objects with radii exceeding 1000 km.  Estimating
stirring timescales for these objects using the approach in this
paper is difficult, because the growth rate of large objects is 
comparable to the stirring timescale in the outer disk (KL99;
\cite{ken99}).  The statistical approach used for these calculations
also begins to break down when the most massive objects are few in
number, although Weidenschilling \etal (1997) note that the stirring
rates for one large body are remarkably close to those derived from 
complete $n$-body calculations (see the Appendix).  Our approach
also does not include orbital resonances, which become important
for planets with masses much larger than those considered here. 
We suspect that the stirring timescales for objects with radii of
1000--3000 km will roughly follow the scaling laws described in 
\S3.3, but cannot confirm this hypothesis without a more 
sophisticated calculation.

We conclude that our stirring times are reasonably accurate estimates
for a debris disk containing a mass distribution of modest-sized
planets and planetesimals with $r \lesssim$ 500 km.  

Future calculations including coagulation and shattering will yield 
predicted surface density distributions for the dust in a model disk.
In the stirring calculations described here,
we adopt an initial surface density distribution $\Sigma \propto r^{-3/2}$ 
which remains fixed with time.  Surface density distributions derived from 
observations range from $\Sigma \propto r^{-1}$ (e.g., Wilner \etal 2000) 
to $\Sigma \propto r^{-3}$ (e.g., Trilling \etal 2000).  Tests with different
surface density distributions suggest that the radial velocity gradient
becomes flatter as the surface density distribution becomes shallower.
We thus expect larger velocities and larger scale heights in disks with
more mass at larger distances from the central star.  The exponent in 
the radial surface density distribution may well evolve with time, because
the timescales to produce large planets in the disk is a strong function
of heliocentric distance (\cite{ken99}).

Our simple models do not explain isolated features observed in debris 
disks such as gaps, warps, and other asymmetries.  Gaps and perhaps
warps are a natural outcome of large planet formation in the disk
(see \cite{art97} and references therein). The size of the gap 
surrounding a large planet with mass $M_P$ is roughly 5--10 Hill radii, 
where $R_H \sim (M_P/3 M_{\star})^{1/3}$
(see the Appendix and references therein).  Large planets can 
cause gaps and rings through mean motion resonances; the size of 
the gap or ring produced by a resonance also depends on $R_H$ 
(e.g., \cite{hol93}).
Bright spots in a disk or ring might be caused by a planet which 
forces eccentric orbits (\cite{wya99}).
More sophisticated treatments of planetary dynamics, such as 
$n$-body simulations, are needed to understand the formation 
and evolution of these features.  The coagulation and gravitational
stirring models described here and in KL99 will eventually yield
plausible initial conditions for $n$-body simulations.

\section{DISCUSSION AND SUMMARY}

On timescales of 10 Myr to 1 Gyr, disks with a size distribution of 
planetesimals can reach conditions needed to produce a collisional 
cascade.
In this model, large bodies stir up small bodies to large velocities;
collisions between smaller bodies initiate the collisional cascade. 
In our calculations with equal mass per mass bin, the disk must be
at least as massive as a minimum mass solar nebula and contain objects
with radii of 500 km or larger.  Larger objects in massive disks stir 
the velocities of small planetesimals more effectively than smaller 
objects in less massive disks.

Once the collisional cascade begins, a simple estimate using the 
coagulation equation indicates that the survival time for the disk 
is roughly 500 Myr for a minimum mass solar nebula with a size of 
100 AU composed primarily of 1 km objects.  More massive disks have 
shorter survival times, because the coagulation timescale is inversely 
proportional to the initial mass in the disk (see KL98 and KL99).
Smaller disks also have shorter survival times.  This estimate is
probably accurate to a factor of 2--3 based on previous, more complete
coagulation calculations in a single annulus (KL99; \cite{ken99}).  
Multi-annulus coagulation calculations now underway will test 
this estimate in more detail.

Our results account for several observed aspects of nearby protostellar
disks.  The estimated disk lifetime of $\sim$ 500 Myr agrees with the
median age of $\sim$ 400 Myr for the central stars of debris disk
systems derived from {\it ISO} and other data (e.g., \cite{hab99}).  
A coagulation model can account for the large range in central star
ages, $\sim$ 10 Myr for HR 4796A up to $\sim$ 1 Gyr for 55 Cnc,
by varying the initial mass of the disk.  In our interpretation,
the disks in young systems such as HR 4796A are initially more massive 
than the disks in older stars such as 55 Cnc.  
Song \etal (2000) note that the median age is larger for less massive
stars with debris disks.  A coagulation model naturally explains
this observation, because the collision time scales with the orbital
timescale, $\tau_{orb} \propto M_{\star}^{-1/2}$.  Future studies with
{\it SOFIA} and {\it SIRTF} should yield larger samples of systems to
test the predicted lifetimes as a function of stellar mass and the
initial mass in the disk.

The observed scale height profiles of debris disk systems place useful 
constraints on coagulation and stirring models.
The predicted scale height distribution $H \propto r^{3/4}$ is shallower
than the typical distribution observed in $\beta$ Pic and other systems
$H \propto r$ (e.g., \cite{art97}, \cite{lag00}).  Radiation
pressure on shattered grains in the inner portions of the disk should
increase the scale height at larger radii; Poynting-Robertson drag on 
shattered grains in the outer parts of the disk should decrease the
scale height at small radii.  Coagulation tends to produce larger 
objects in the inner portions of the disk and should produce larger
scale heights in the inner disk.  Although more detailed calculations
are needed to see how these competing physical processes act on the
observed scale height, we are encouraged that the scale height 
distribution derived solely from the stirring calculations is close 
to those observed in real debris disk systems.

We acknowledge a generous allotment of computer time on the HP Exemplar 
`Neptune' and the Silicon Graphics Origin-2000 `Alhena' through funding 
from the NASA Offices of Mission to Planet Earth, Aeronautics, and Space 
Science.

\vfill
\eject

\appendix

\section{APPENDIX}

\subsection{Overview}

We assume that planetesimals are a statistical ensemble of masses 
in $N$ concentric, cylindrical annuli with width $\Delta a_i$,
and height $H_i$ centered at radii $a_i$ from a star with mass
$M_{\star}$ and luminosity $L_{\star}$.  
The particles have horizontal $h_{ik}(t)$ and vertical $v_{ik}(t)$
velocity dispersions relative to an orbit with mean Keplerian velocity 
$V_{Ki}$ (see \cite{lis93}).  We approximate the continuous distribution 
of particle masses with discrete batches having an integral number of 
particles $n_{ik}(t)$ and total masses $M_{ik}(t)$.  The average mass 
of each of $M$ mass batches, $m_{ik}(t)$ = $M_{ik}(t) / n_k(t)$, evolves 
with time as collisions add and remove bodies from the batch (\cite{wet93}).  

KL98 and KL99 describe our approach for solving the evolution of 
particle numbers and velocities for a mass batch $k$ in a single
annulus $i$ during a time step $\delta t$.  Here we generalize the
velocity evolution from elastic collisions for a set of annuli.  
Interactions occur between particles with masses $m_{ik}$ in annulus
$i$ and $m_{jl}$ in annulus $j$.  Each annulus has a geometric width
$\Delta a_i$.  Annuli with a fixed constant size have 
$\Delta a_i = a_{i+1} - a_i$. Annuli with a variable width 
have $\Delta a_i = \Delta a_0 \cdot a_i$ and
$a_i = a_{in} ((1 + 0.5 \Delta a_0)/(1 - 0.5 \Delta a_0))^i$,
where $a_{in}$ is the inner edge of the first annulus.
Following Weidenschilling \etal (1997), particles interact if their 
orbits approach within 2.4 times their mutual Hill radius $R_H$. 
The `overlap region' for elastic collisions is 

\begin{equation}
o_{ijkl,el} = 2.4 R_H + 0.5 (w_{ik} + w_{jl}) - | a_i - a_j | ~ ;
\end{equation}

\noindent
where $w_{ik}$ is the radial extent of the orbit of particle
$k$ with orbital eccentricity $e_k$ in annulus $i$:

\begin{equation}
w_{ik} = \left\{ \begin{array}{l l l}
         \Delta a_i + e_k a_i & \hspace{5mm} & e_k a_i \le \Delta a_i \\
         (\Delta a_i + e_k a_i) (e_k a_i / \Delta a_i)^{1/4} & \hspace{5mm} & e_k a_i > \Delta a_i \\
 \end{array}
         \right .
\end{equation}

\subsection{Velocity Evolution}

We solve a set of Fokker-Planck equations to follow the time-evolution
of $h_{ik}$ and $v_{ik}$ (\cite{hor85}; \cite{wet93}; \cite{ste00}):

\begin{equation}
\frac{dh_{vs,ik}^2}{dt} = \sum_{j=1}^{j=N} \sum_{l=1}^{l=M} f_{ij} C~(h_{ik}^2 + h_{jl}^2)~m_{jl}~ B_{\Lambda} J_e(\beta)
\end{equation}

\begin{equation}
\frac{dv_{vs,ik}^2}{dt} = \sum_{j=1}^{j=N}  \sum_{l=1}^{l=M} f_{ij} \frac{C}{\beta_{kl}^2}~(v_{ik}^2 + v_{jl}^2)~m_{jl}~B_{\Lambda}~J_z(\beta)
\end{equation}

\noindent
for viscous stirring and

\begin{equation}
\frac{dh_{df,ik}^2}{dt} = \sum_{j=1}^{j=N} \sum_{l=1}^{l=M} f_{ij} C~(m_{jl}h_{jl}^2 - m_{ik}h_{ik}^2)~A_{\Lambda}~H_e(\beta)
\end{equation}

\begin{equation}
\frac{dv_{df,ik}^2}{dt} = \sum_{j=1}^{j=N} \sum_{l=1}^{l=M} f_{ij} \frac{C}{\beta_{kl}^2}~(m_{jl}v_{jl}^2 - m_{ik}v_{ik}^2)~A_{\Lambda}~H_z(\beta)
\end{equation}

\noindent
for dynamical friction.  In these expressions, $N$ is the number of
annuli and $M$ is the number of mass batches within each annulus;
$\beta_{kl}^2 = (i_{ik}^2 + i_{jl}^2)/(e_{ik}^2 + e_{jl}^2)$, and
$C = G^2 \rho_l / (\sqrt{\pi} V_K^3(h_{ik}^2 + h_{jl}^2)^{3/2})$ 
is a function of the density of particles in batch $l$ and the 
relative horizontal velocity of the mass batches (\cite{wet93}, 
\cite{ste00}). The functions $H_e$, $H_z$, $J_e$, and $J_z$ are 
definite integrals defined in Stewart \& Ida (2000).  
The overlap fraction $f_{ij}$ is the fraction of bodies in annulus
$i$ that approach within 2.4 $R_H$ of the bodies in annulus $j$.
We set $\rho_l = M_l/V_l$, where $M_l$ is the total mass of bodies
with $m_l$ in annulus $j$ and $V_l$ is the larger of the two volumes
for the interacting annuli.  Following Stewart \& Ida (2000), we 
also set $A_{\Lambda} = {\rm ln} (\Lambda^2 + 1) $ and 
$B_{\Lambda} = A_{\Lambda} - \Lambda^2 / (\Lambda^2 - 1)$,
where 

\begin{equation}
\Lambda = \left ( \frac{m_{ik} + m_{jl}}{M_{\star}} \right ) ^{-1} \left ( \frac{H_{ijkl}}{a_{avg}} + \frac{R_H}{a_{avg}} \right ) ~ V_{ij}^2 ~ ,
\end{equation}

\noindent
is for calculations without physical collisions and

\begin{equation}
\Lambda = \left ( \frac{H_{ijkl} + G (m_{ik} + m_{jl}) V_{ij}^{-2}}{(r_{ik} + r_{jl}) (1 + 2 (V_{e,ij}/V_{ij})^2} \right ) ~ ,
\end{equation}

\noindent
is for calculations with physical collisions.
In these expressions,
$G$ is the gravitational constant,
$H_{ijkl}$ is the mutual scale height, 
$a_{avg} = 0.5 (a_i + a_j)$, and
$R_H$ is the mutual Hill radius,
$R_H = ((m_{ik} + m_{jl})/3 ~ M_{\star})^{1/3} a_{avg}$.

This formulation for velocity evolution
breaks down in the low velocity regime, when the relative velocities 
of particles approach the 
Hill velocity, $v_H = R_H V_K / a $. KL99 described a simple 
solution to this problem, based on a comparison of $N$-body integrations 
(\cite{ida90}) with published scaling laws for velocity evolution 
at low velocities (\cite{bar91}; \cite{wet93}; \cite{wei97}).  This
solution fails at very low velocities. We thus replaced equations 
(A20) and (A21) of KL99 with the more accurate results of Ida \& Makino
(1992, 1993):

\begin{equation}
\left ( \frac{dh_{vs,ik}^2}{dt} \right )_{lv} = \sum_{j=1}^{j=N} \sum_{l=1}^{l=M} C_{h,vs}^{\prime} \left (
\frac{G^2}{m_{ik} + m_{jl}} \right )^{2/3} \rho_{jl}~m_{jl}~H_{jl}~\Omega_{ij}^{1/3}
\end{equation} 

\begin{equation}
\left ( \frac{dv_{vs,ik}^2}{dt} \right )_{lv} = \sum_{j=1}^{j=N} \sum_{l=1}^{l=M} C_{i,vs}^{\prime} \left (
\frac{G}{(m_{ik} + m_{jl})^2} \right )^{2/3} \Omega_{ij}^{-1/3}~\rho_{jl}~m_{jl}~H_{jl} (v_{ik}^2 + v_{jl}^2)
\end{equation} 

\vskip 2ex
\noindent
for viscous stirring and

\begin{equation}
\left ( \frac{dh_{df,ik}^2}{dt} \right )_{lv} = \sum_{j=1}^{j=N} \sum_{l=1}^{l=M} C_{e,df}^{\prime} \left (
\frac{\Omega_{ij} R_H^2 \rho_{jl} H_{ijkl}}{m_{ik} m_{jl}} \right ) \left ( m_{jl} h_{jl}^2  - m_{ik} h_{ik}^2 \right )
\end{equation} 

\begin{equation}
\left ( \frac{dv_{df,ik}^2}{dt} \right )_{lv} = \sum_{j=1}^{j=N} \sum_{l=1}^{l=M} C_{i,df}^{\prime} \left (
\frac{\Omega_{ij} R_H^2 \rho_{jl} H_{ijkl}}{m_{ik} m_{jl}} \right ) \left ( m_{jl} v_{jl}^2  - m_{ik} v_{ik}^2 \right )
\end{equation} 

\noindent
for dynamical friction.
In these expressions, 
$\Omega_{ij}$ is the average angular frequency of the two Keplerian orbits,
and $H_{jl}$ is the vertical scale height of particles in annulus $l$.
The $C$'s are constants:
$C_{h,vs}^{\prime}$ = 10,
$C_{v,vs}^{\prime}$ = 1, and
$C_{h,df}^{\prime}$ = $C_{v,df}^{\prime}$ = 5.
We adjusted the values of these to agree with results of $n$-body 
calculations described below.

The Fokker-Planck equations -- and their low velocity substitutes --
do not include stirring rates for distant perturbations of particles 
whose orbits do not cross.  Weidenschilling (1989) and 
Hasegawa \& Nakazawa (1990) derived expressions for the rate of
change of orbital eccentricities and inclinations for distant
encounters.  Stewart \& Ida (2000) show that these perturbation
calculations underestimate the strength of distant encounters 
when the orbital separation is smaller than several $R_H$. They
derive expressions appropriate for velocity evolution in a single
annulus and show that their results agree with $n$-body simulations.
We follow Weidenschilling \etal (1997) and use the stirring rates 
defined in Weidenschilling (1989).  These are:

\begin{equation}
\frac{dh_{lr,ik}^2}{dt} = \sum_{j=1}^{j=N} \sum_{l=1}^{l=M} C_{lr,e} x_{ij} \frac{G^2 \rho_{jl} M_{jl}}{\Omega_{avg}} \left ( \frac{{\rm tan^{-1}}(H_{ijkl}/D_{min})}{D_{min}} - \frac{{\rm tan^{-1}}(H_{ijkl}/D_{max})}{D_{max}} \right )
\end{equation}

\noindent
for continuum bodies and

\begin{equation} 
\frac{dh_{lr,ik}^2}{dt} = \sum_{j=1}^{j=N} \sum_{l=1}^{l=M} \frac{G^2}{\pi \Omega a} \left ( \frac{C_{lr,e}^{\prime} m_l^2}{(\delta a^2 + 0.5 H_{jl}^2)^2} \right )
\end{equation}

\noindent
where $D_{min} = {\rm max} (2.4 R_H, 1.6(h_{ik}^2 + h_{jl}^2)^{1/2})$,
$D_{max} = 0.5 ~ {\rm max} (w_{ik}, w_{jl})$, and
$\delta a$ = $ | a_i - a_j | $.  We do not calculate long-range 
stirring rates for inclination, because these rates are much smaller
than rates for the eccentricity (Weidenschilling 1989; Stewart \& Ida 2000)
These equations,
together with equations (A9)--(A12), yield satisfactory agreement
between our calculations and the results of more accurate direct
$n$-body integrations for 
$C_{lr,e}$ = 23.5 and
$C_{lr,e}^{\prime}$ = 5.9
(see below).  Our choices for the constants in these expressions 
are larger than those of Weidenschilling (1989) to mimic the
increased long-range stirring derived by Stewart \& Ida (2000).

To solve the set of coagulation and velocity evolution equations,
we employ a fourth order Runge-Kutta solution (\cite{pre92}).
For each timestep, we derive separate solutions for a full timestep of 
length $\delta t$ and for two half timesteps of length $\delta t$/2.
These solutions converge if the maximum difference between the dynamical
variables is $\delta_{err} < \delta_{max}$; we adopt $\delta_{max} = 
10^{-5}$ for most calculations.  We follow Press \etal (1992) and 
increase the timestep by $(\delta_{err}/\delta_{max})^{0.20}$ when the
solution converges and decrease the timestep by 
$(\delta_{err}/\delta_{max})^{0.25}$ when the solution does not converge.

Our algorithm evolves quantities which lie in bins of radius, a
discretization which provides a good starting point for parallel
computation. We use explicit message-passing functions to distribute 
the summations required in the evolution equations to multiple
processors. Each processor handles a unique set of radial bins
and performs summations only for those indices. The computations
depend on access to information from all radial bins, hence prior 
to each time step, we communicate the state of quantities in all
bins to all processors.  For the number of processors we use, up to 32
on an HP Exemplar and an SGI Origin~2000, this communication load is
not great and the algorithm speeds up nearly linearly with the number
of processors.

\subsection{Tests of the Evolution Code}

To test the velocity evolution algorithm, we attempt to reproduce
published $n$-body and particle-in-a-box calculations. Stewart \&
Ida (2000), Ida \& Makino (1993), and Kokubo \& Ida (1995) describe 
$n$-body
calculations of gravitational stirring of planetesimals in orbit
at 1 AU; Weidenschilling \etal (1997) and Stewart \& Ida (2000) show 
that particle-in-a-box calculations roughly match the $n$-body results.

We begin with tests of the `standard' velocity evolution algorithm
in the high velocity limit.  Stewart \& Ida (2000) describe several
comparisons of stirring calculations in a single annulus with detailed
$n$-body calculations.  Here we try to reproduce the $n$-body results
using a collection of $N$ annuli.  The first test considers 800 
planetesimals with masses $m_1 = 10^{24}$ g orbiting the Sun at 
$a_0$ = 1 AU.  These bodies are evenly distributed on a grid with 
inner radius $a_{in} = a_0 - 40 R_H$ and outer radius $a_{out} = 
a_0 + 40 R_H$; $R_H$ is the Hill radius for collisions between 
these bodies, $R_H$ = 0.0007 AU.  We consider grids with 20, 40,
and 80 annuli to test the sensitivity of the results to the
grid spacing.  The surface density is $\Sigma$ = 10 g cm$^{-2}$. 
The bodies have $e_0 = 2 i_0 = $ $2 \times 10^{-3}$.  

Figure 9 shows that the evolution of $e$ and $i$ is independent
of the grid spacing.  We plot $e$ and $i$ in units of the Hill
radius for the central four annuli of the model grid.  
Our results are nearly identical to the $n$-body 
calculations of Stewart \& Ida (2000).  At 20,000 yr, our result
for $e$ is $\sim$ 5\% smaller than the Stewart \& Ida result; 
our value for $i$ is indistinguishable from the Stewart \& Ida 
result.  The small difference in the eccentricity evolution probably 
results from long range stirring; Stewart \& Ida (2000) use another 
algorithm for long-range stirring which is more computationally 
expensive and probably more accurate than the one we use here.  

Figure 10 shows the radial profile of the velocity distribution 
at three points in the evolution.  The velocity profiles become 
more bow-shaped at late times, because objects at the edges of 
the grid have fewer neighbors than objects in the middle of the
grid.  The shape of this bow is independent of the grid spacing,
as indicated by the dot-dashed and dashed lines in Figure 10.

Figure 11 compares the results of the first test with two
other tests described  by Stewart \& Ida (2000). These tests
follow the evolution of (a) 8000 planetesimals with $m_1 = $
$10^{24}$ g with a surface density of 100 g cm$^{-2}$ 
and (b) 400 planetesimals with $m_1 = $ $10^{26}$ g and 
a surface density of 500 g cm$^{-2}$.
For another comparison, we consider (c) $4 \times 10^5$
planetesimals with $m_1 = $ $10^{23}$ g and the same surface density
as (b). Stewart \& Ida (2000) demonstrated that the time
evolution of these tests scales with the mass of a planetesimal
and the surface density as

\begin{equation}
\frac{de}{dt} \propto \left ( \frac{\sigma}{\rm 10 ~ g ~ cm^{-2}} \right ) \left ( \frac{10^{24} \rm ~ g}{m_1} \right )^{1/3} ~ \frac{de_H}{dt^{\prime}}
\end{equation}

\begin{equation}
\frac{di}{dt} \propto \left ( \frac{\sigma}{\rm 10 ~ g ~ cm^{-2}} \right ) \left ( \frac{10^{24} \rm ~ g}{m_1} \right )^{1/3} ~ \frac{di_H}{dt^{\prime}} ~ ,
\end{equation}

\noindent
where $t^{\prime}$ is a scaled time.
The dashed and dot-dashed curves in Figure 11 show the scaled 
evolution for tests (a), (b), and (c).  There is excellent agreement  
between these results and the time evolution of the eccentricity
and inclination shown in Figure 9: the models with $m_1 = 10^{23}$ g
lag test (a) by $\sim$ 1\% throughout the evolution, while
models with $m_1 = 10^{26}$ g run ahead by $\sim$ 1\%.  

To test the low velocity limit of the multi-annulus code, we consider
a calculation of $N = 805$ planetesimals with masses $m_1 = $
$2 \times 10^{24}$ g orbiting the Sun at $a_0$ = 1 AU. The 
planetesimals are evenly distributed on a grid with inner radius 
$a_{in} = a_0 - 17.5 R_H$ and outer radius $a_{out} = a_0 + 17.5 R_H$;
$R_H$ is the Hill radius for collisions between $m_1$ and
$m_2 = 2 \times 10^{26}$ g, $R_H$ = 0.003235 AU.  The 35 annuli in
this grid are spaced at intervals of $R_H$. The surface density is 
$\Sigma$ = 10 g cm$^{-2}$. The bodies have $e_0 = i_0 = 3.235$ 
$\times 10^{-5}$.  

Figure 12 shows the time evolution of $(e^2 + i^2)$ for these bodies 
in units of the Hill radius.  The velocities of all planetesimals 
increase uniformly with time; the average eccentricity is 
$e \approx 0.7 R_H$ at $t = 100$ yr,
$e \approx 1.27 R_H$ at $t = 800$ yr, and
$e \approx 1.63 R_H$ at $t = 2000$ yr.  Particles at the edge of 
the grid have somewhat smaller velocities, because they do not 
have as many neighbors.  The dashed line in the Figure plots the 
eccentricity evolution for a model grid with 70 annuli spaced 
at intervals of $R_H$; the difference between this model and the 
`standard' grid with 35 annuli is small.

Figure 13 shows how the $2 \times 10^{24}$ g planetesimals react 
when a single object with $m_2 = 2 \times 10^{26}$ g lies at $a_0$.  
Velocities of small planetesimals within $\pm$3--4 $R_H$ of the 
larger planetesimal increase rapidly due to short range stirring.  
Velocities of more distant planetesimals increase slowly due to 
long range stirring by the large planetesimal and short range 
stirring by nearby small planetesimals.  The radial profile of the
velocity distribution for the central 9 annuli is fairly flat at
$t$ = 2000 yr, and has a similar amplitude and shape as the velocity
distribution for an $n$-body calculation with identical starting
parameters (see Ida \& Makino 1993).  Our calculation yields fewer
annuli in the central high velocity peak compared to the $n$-body 
calculation.  Planetesimals 
in the center of the $n$-body grid migrate away from the massive 
object and then stir up planetesimals farther out in the grid; our
algorithm does not include this evolution.  Weidenschilling \etal 
(1997) obtained results similar to ours.

Figures 14 and 15 show the evolution of the planetesimal swarm 
when another large planetesimal is added.  The two large planetesimals
at $\pm 5 R_H$ in Figure 14 initially interact only by long range 
stirring; the early evolution of small planetesimals near each of
the large planetesimals thus follows the evolution shown in Figure 13.
Once the orbits near the two large bodies begin to cross, the
velocities of the smaller planetesimals increase more rapidly than
those near a single large body.  The velocity profile is still flat
at $t$ = 2000 yr; the area of the high velocity peak is roughly twice
the area of the high velocity peak in Figure 13.

The two large planetesimals at $\pm 2 R_H$ in Figure 15 interact
by short range stirring throughout the evolution.  Small planetesimals
rapidly increase in velocity and reach $(e^2 + i^2)^{1/2}$
$\approx$ 5.5 $R_H$ in 2000 yr.  The velocities of the large 
planetesimals increase as well and approach $(e^2 + i^2)^{1/2}$
$\approx$ 2 $R_H$ at the end of the test.  Our results for the
small planetesimals agree well with the $n$-body results of 
Kokuba \& Ida (1995) and the multizone calculations of 
Weidenschilling \etal (1997).  The evolution of the larger bodies
in our calculations is also similar to those of other calculations
for large initial separations (as in Figure 14).  We do not treat
short-range interactions as well as other calculations that follow
the actual orbits of larger bodies. 

Finally, Figure 16 shows the results of a calculations with 8 large
planetesimals, $m_2 = 2 \times 10^{26}$ g, embedded in a uniform
sea of 805 smaller bodies with $m_1 = 2 \times 10^{24}$ g.  As in
Kokubo \& Ida (1995) and Weidenschilling \etal (1997), the small bodies
achieve large velocities, $(e^2 + i^2)^{1/2}$ $\approx$ 6.5 $R_H$ in 2000 yr.
The shape and amplitude of our radial distribution of $(e^2 + i^2)^{1/2}$
agrees well with previous results.  The velocities of the large bodies
increase uniformly as well and reach $(e^2 + i^2)^{1/2}$ $\approx$ 3 $R_H$
at the end of the evolution.  The velocities of the large bodies in
Figure 16 agree with the median velocity of bodies in the Kokubo \& Ida 
(1995) and the Weidenschilling \etal (1997) calculations.  Our calculations
do not display the same chaotic behavior as these more detailed calculations,
because we do not follow individual orbits for the large bodies.

We conclude that our multi-annulus planetesimal evolution code matches
a variety of tests reasonably well.  The code successfully follows the
evolution of large numbers of interacting planetesimals with a range of
masses.  The code also successfully calculates the velocity evolution of
small numbers of large planetesimals when close interactions are unimportant.
Our code does not account for the chaotic behavior of these close encounters,
but it does yield reasonably good velocities averaged over several close
encounters.  We plan to consider better treatments of these `one-on-one'
interactions in a future study.

\vfill
\eject

\clearpage


\epsfxsize=8.0in
\hskip -10ex
\epsffile{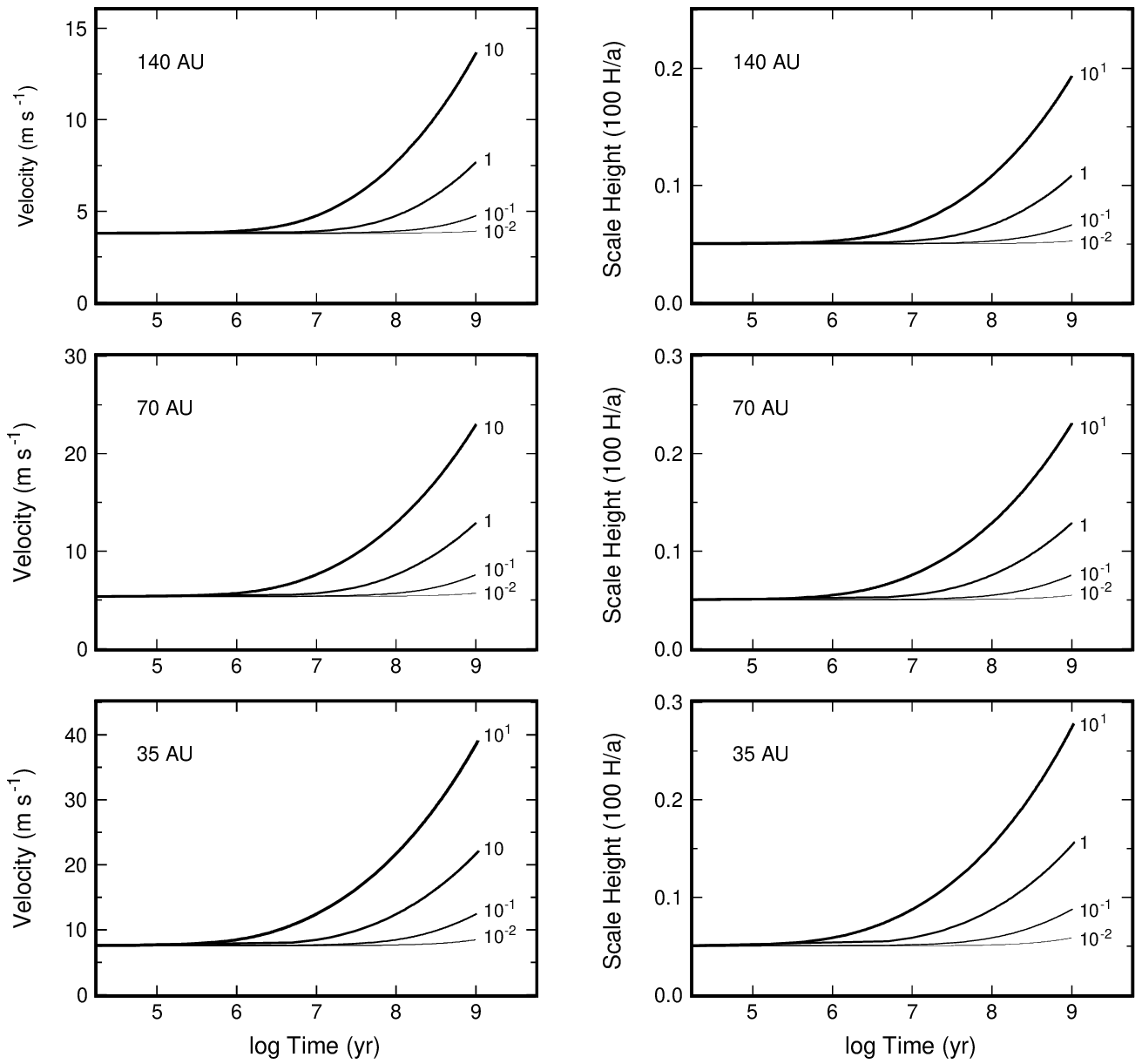}

\figcaption[Kenyon.fig1.ps]
{Evolution of particle velocity (left panels) and vertical scale
height (right panels) for 10 km planetesimals in orbit around a
3 \msun~star at 35 AU (lower panels), 70 AU (middle panels),
and 140 AU (top panels).  The four curves in each panel show
the evolution for planetesimals with normalized surface densities,
$x$, indicated to the right of each curve. Models with $x$ = 1
have a surface density of solids comparable to that in the minimum
mass solar nebula.}

\epsfxsize=8.0in
\hskip -10ex
\epsffile{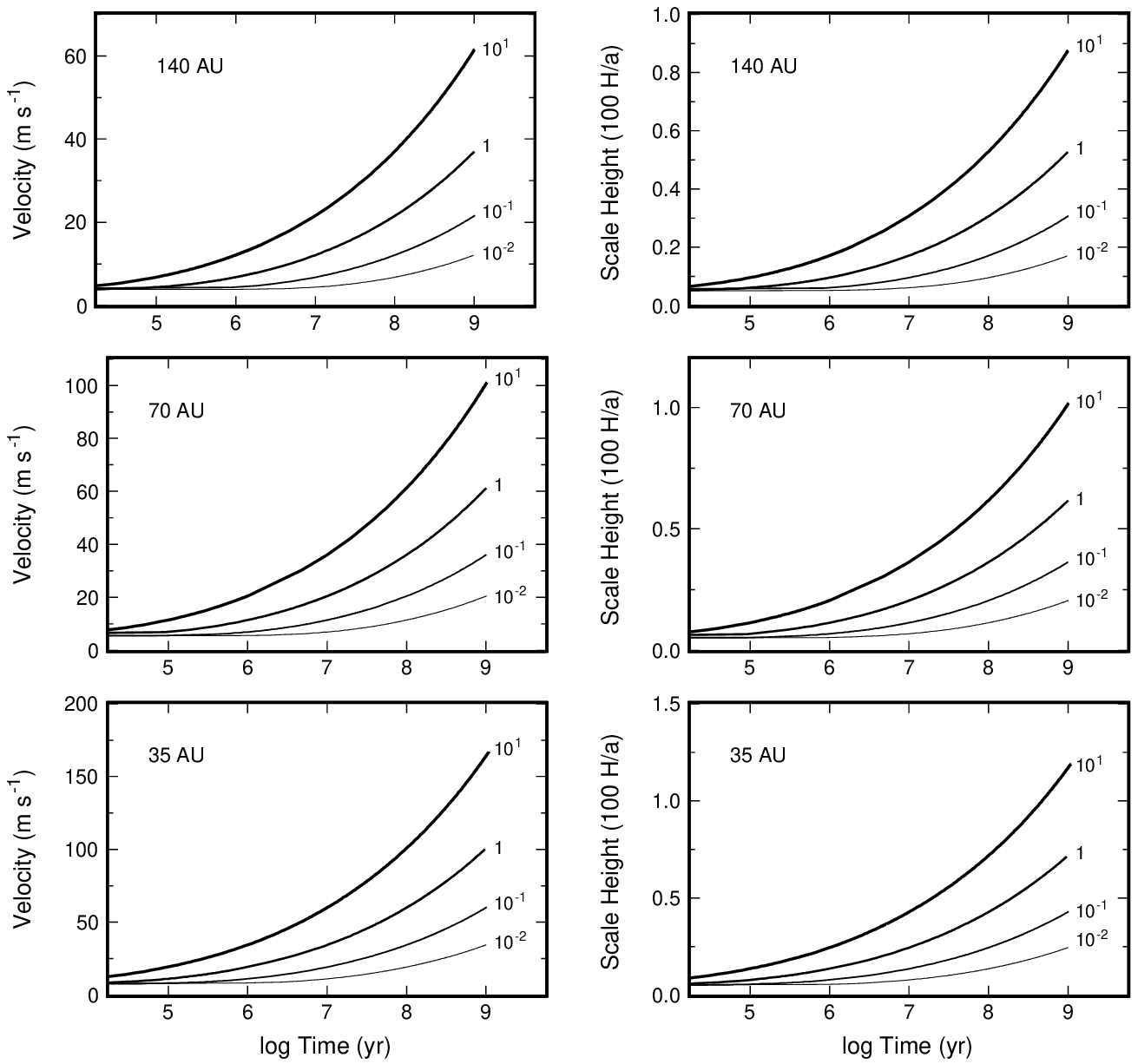}

\figcaption[Kenyon.fig2.ps]
{As in Figure 1, for 100 km planetesimals.}

\epsfxsize=8.0in
\epsffile{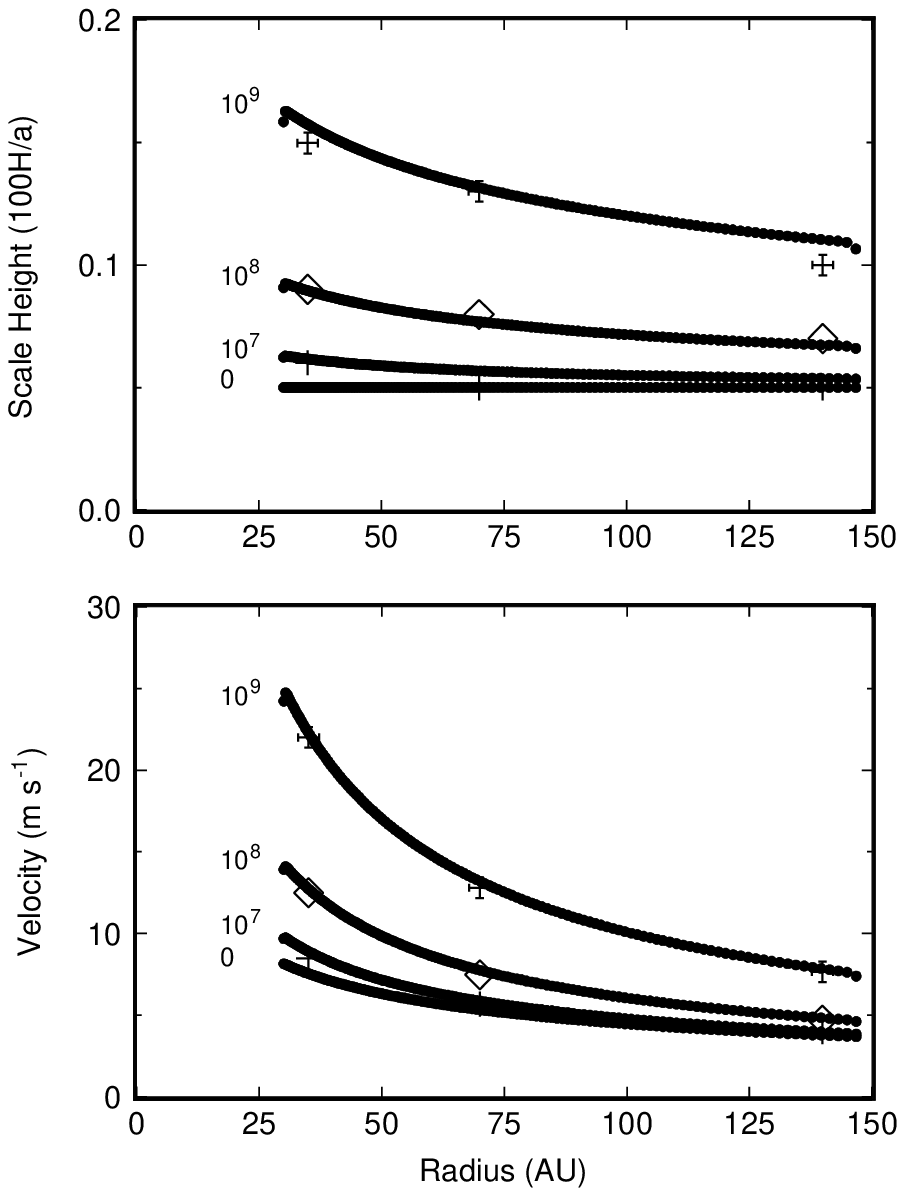}

\figcaption[Kenyon.fig3.ps]
{Evolution of particle velocity (lower panel) and vertical scale  
height (top panel) for 10 km planetesimals in a disk surrounding
a 3 \msun~star.  The initial surface density in the disk is
$\Sigma = 60 (a/{\rm 1 ~ AU})^{-3/2}$.  The evolution time in
years is listed to the left of each curve.}

\epsfxsize=8.0in
\epsffile{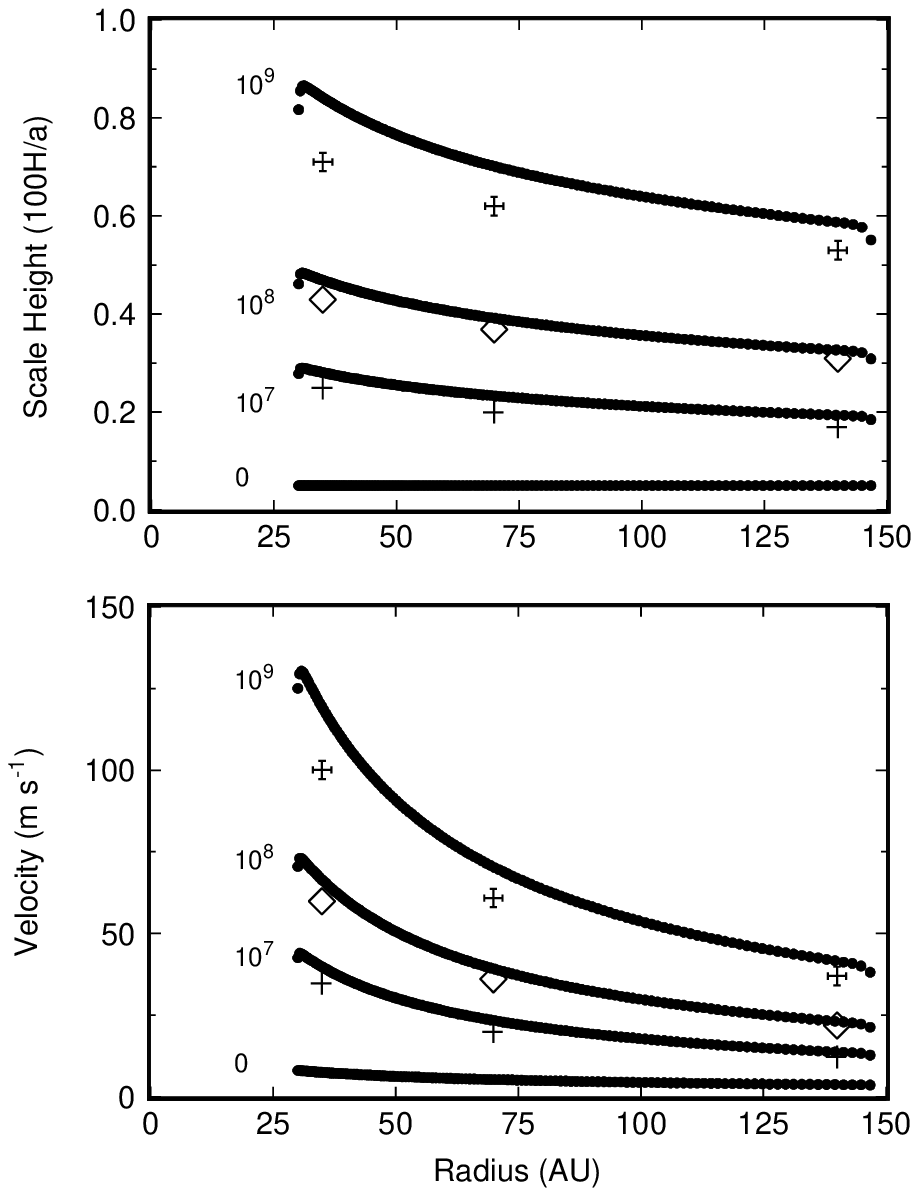}

\figcaption[Kenyon.fig4.ps]
{As in Figure 3, for 100 km planetesimals.}

\epsfxsize=8.0in
\hskip -10ex
\epsffile{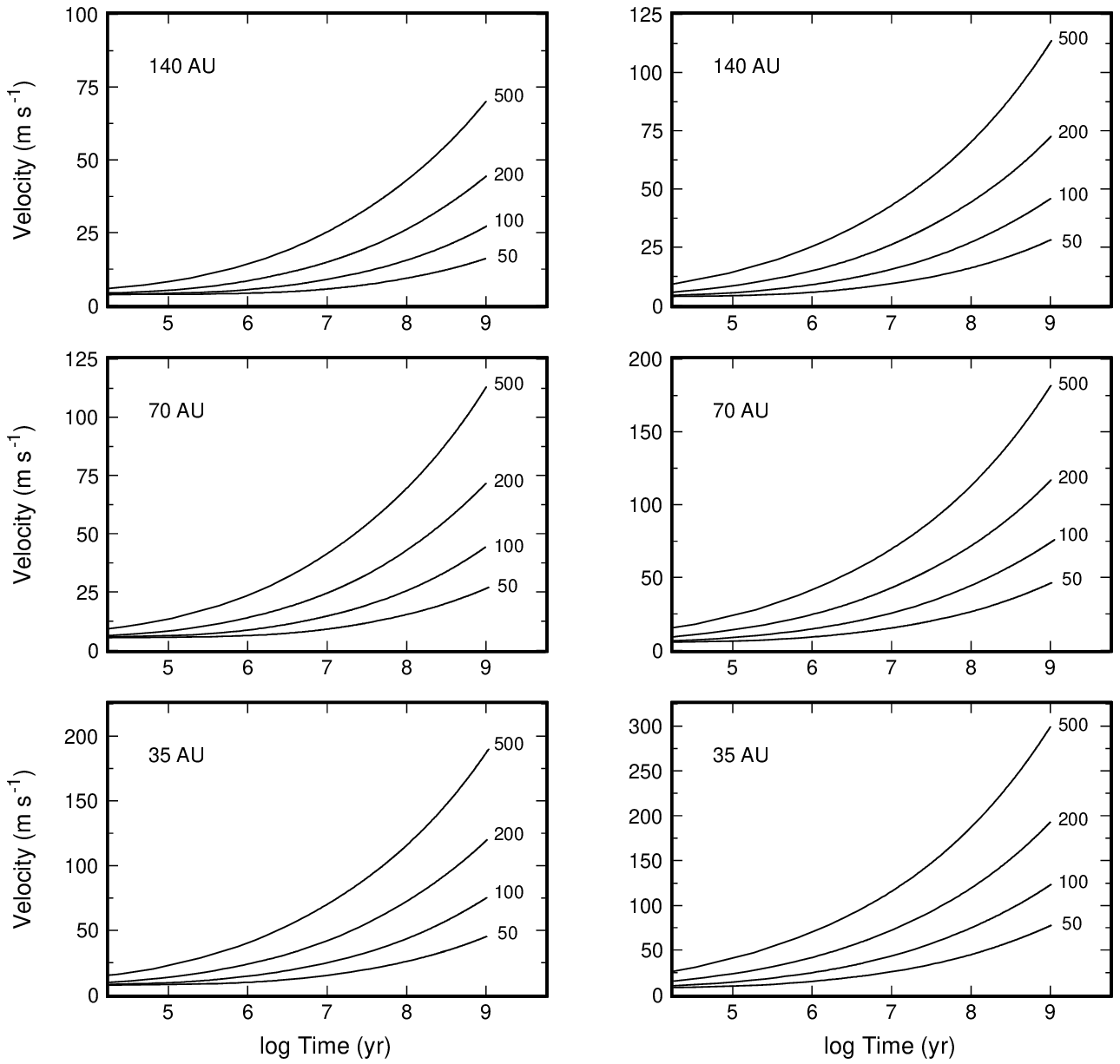}

\figcaption[Kenyon.fig5.ps]
{Evolution of particle velocity for a size distribution of planetesimals 
in orbit around a 3 \msun~star at 35 AU (lower panels), 70 AU (middle 
panels), and 140 AU (top panels).  The size distributions have equal
mass per mass batch; the mass spacing factor is $\delta = 10^3$.  The 
left panels show velocities for small planetesimals, with radii 
$\lesssim$ 1 km, for disks with surface densities equivalent to the 
minimum mass solar nebula; the right panels show results for disks 
with 10 times the mass in a minimum mass solar nebula. The numbers to 
the right of each curve list the maximum size of the planetesimal in
each calculation.  Large planetesimals stir up disks more rapidly
than small planetesimals.}

\epsfxsize=8.0in
\epsffile{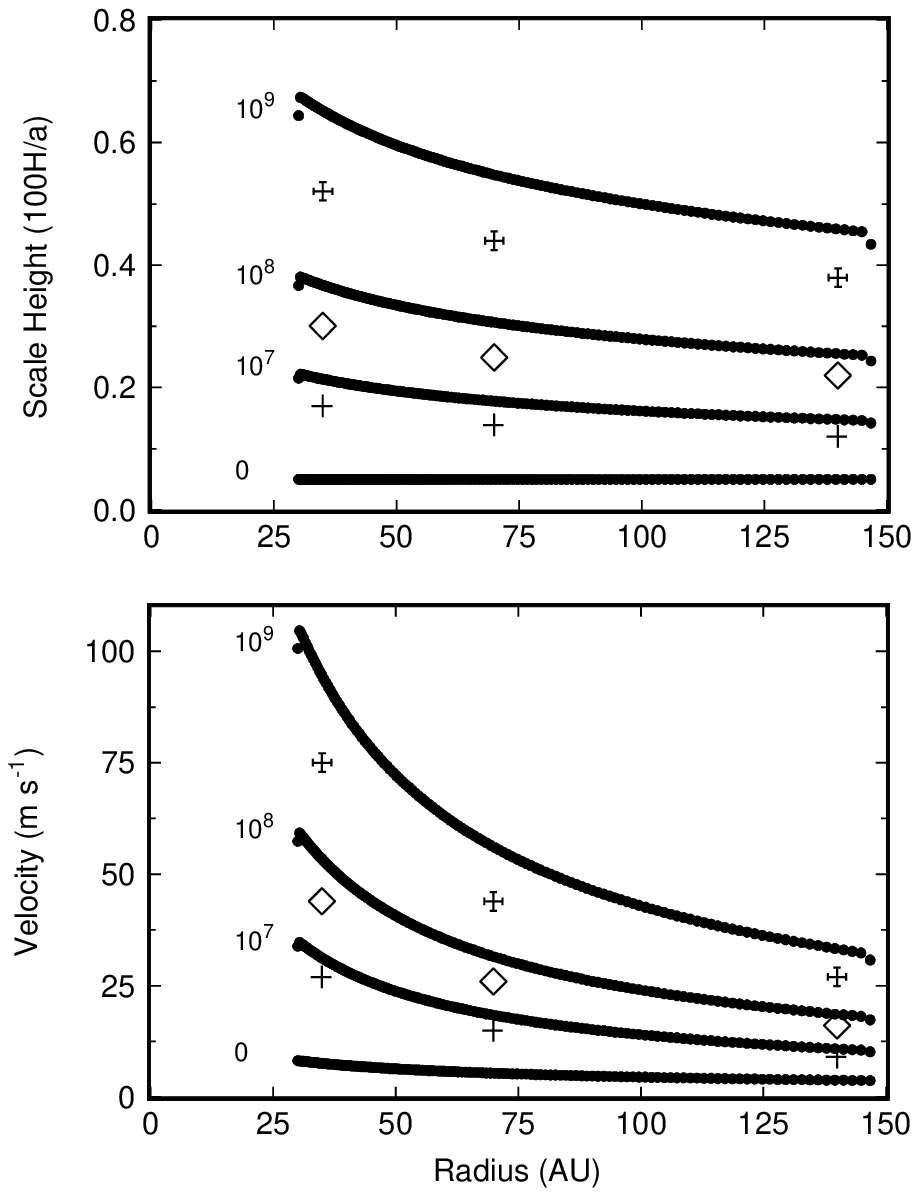}

\figcaption[Kenyon.fig6.ps]
{Evolution of particle velocity (lower panel) and vertical scale
height (top panel) for small planetesimals of a size distribution 
of 1 m to 100 km planetesimals in a disk surrounding a 3 \msun~star.  
The initial surface density in the disk is $\Sigma = 
60 (a/{\rm 1 ~ AU})^{-3/2}$.  The evolution time in years is 
listed to the left of each curve.}
 
\epsfxsize=8.0in
\epsffile{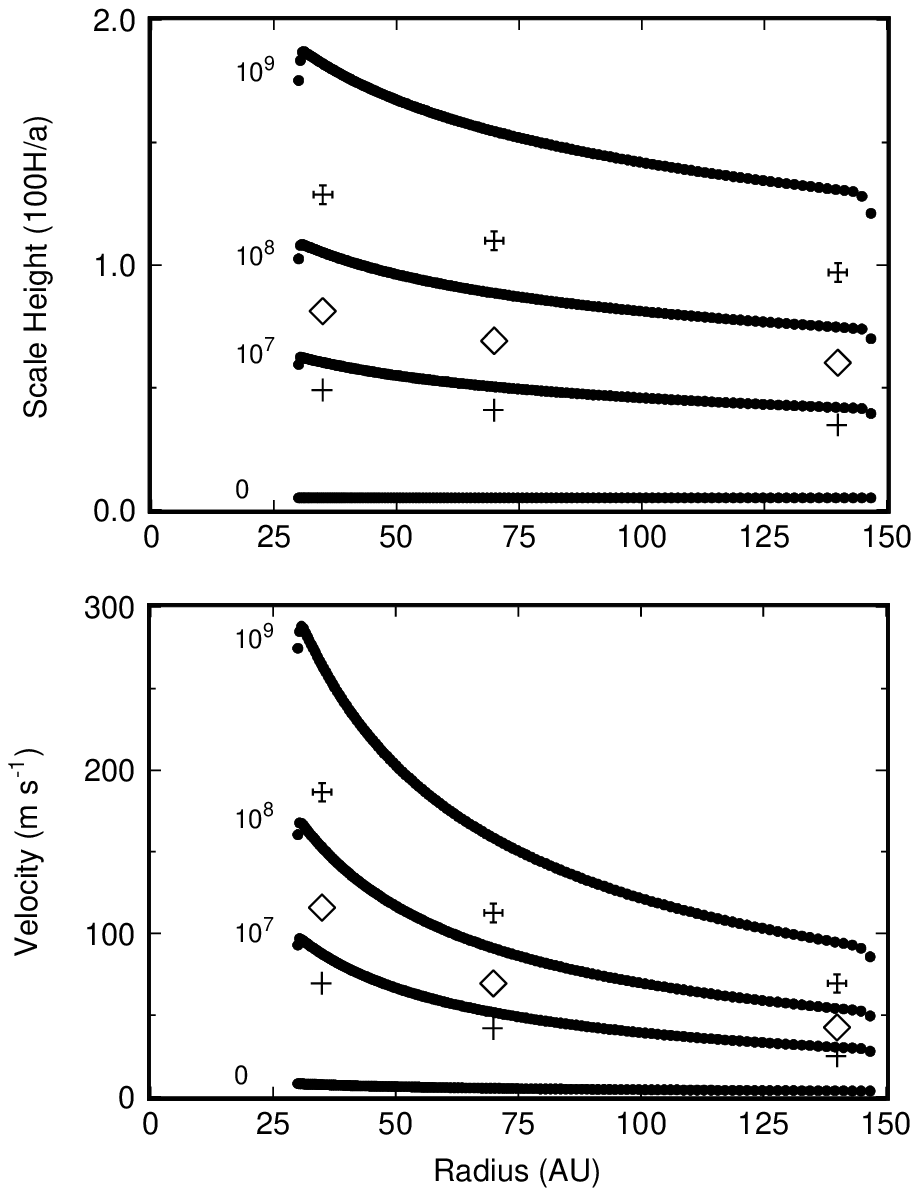}

\figcaption[Kenyon.fig7.ps]
{As in Figure 6, for a size distribution of 1 m to 500 km planetesimals.}

\hskip -5ex
\epsffile{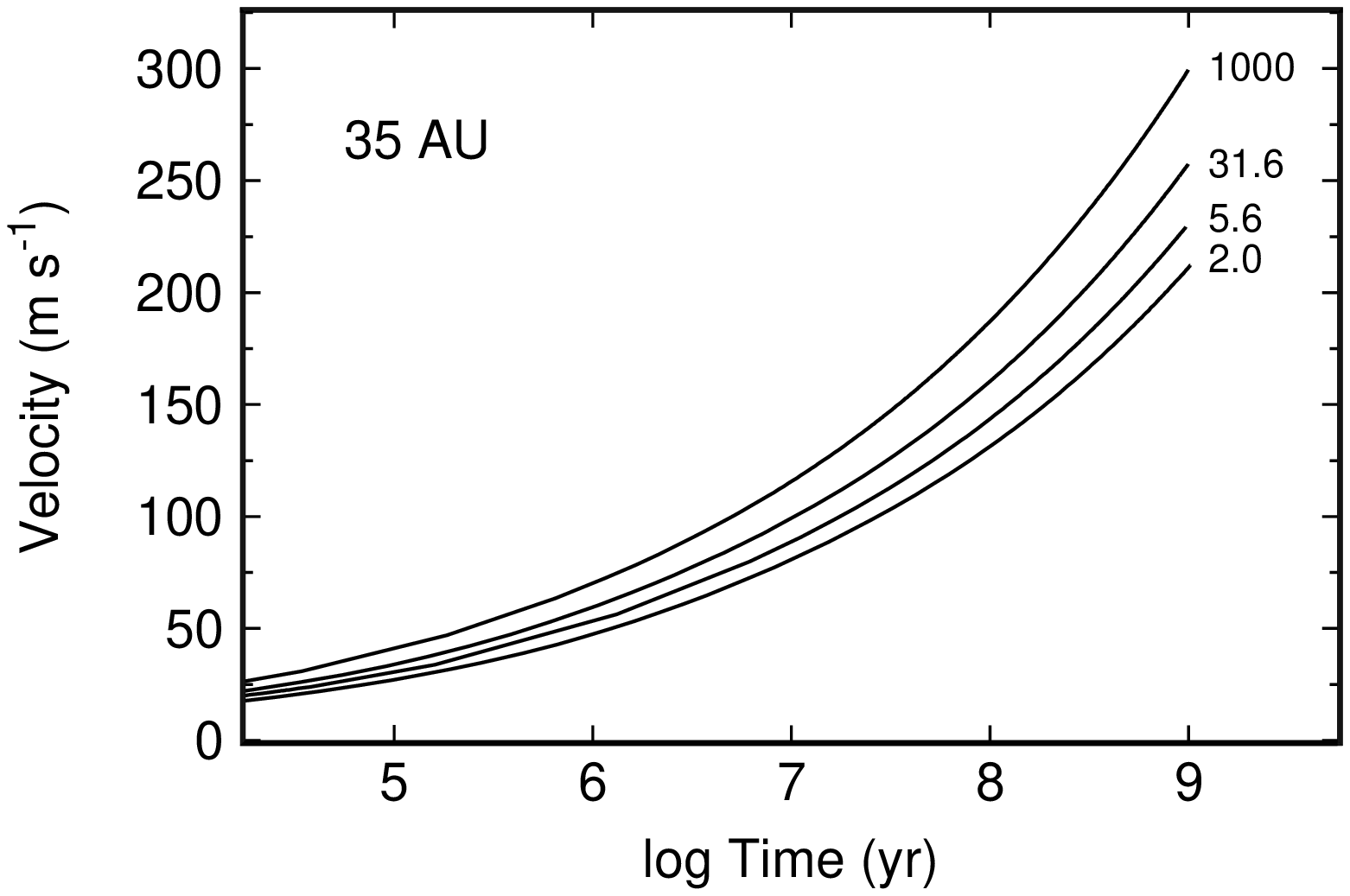}

\figcaption[Kenyon.fig8.ps]
{Evolution of particle velocity at 35 AU as a function of mass spacing
factor, $\delta$, indicated to the right of each curve.  The curves
show the variation of particle velocity for small planetesimals, with 
radii of 1 m to 1 km, in a size distribution with a maximum radius of 
500 km and in a disk with surface density of 10 times the minimum mass
solar nebula.  Models with higher mass resolution achieve lower velocities
compared to models with lower mass resolution.}

\hskip 10ex
\epsffile{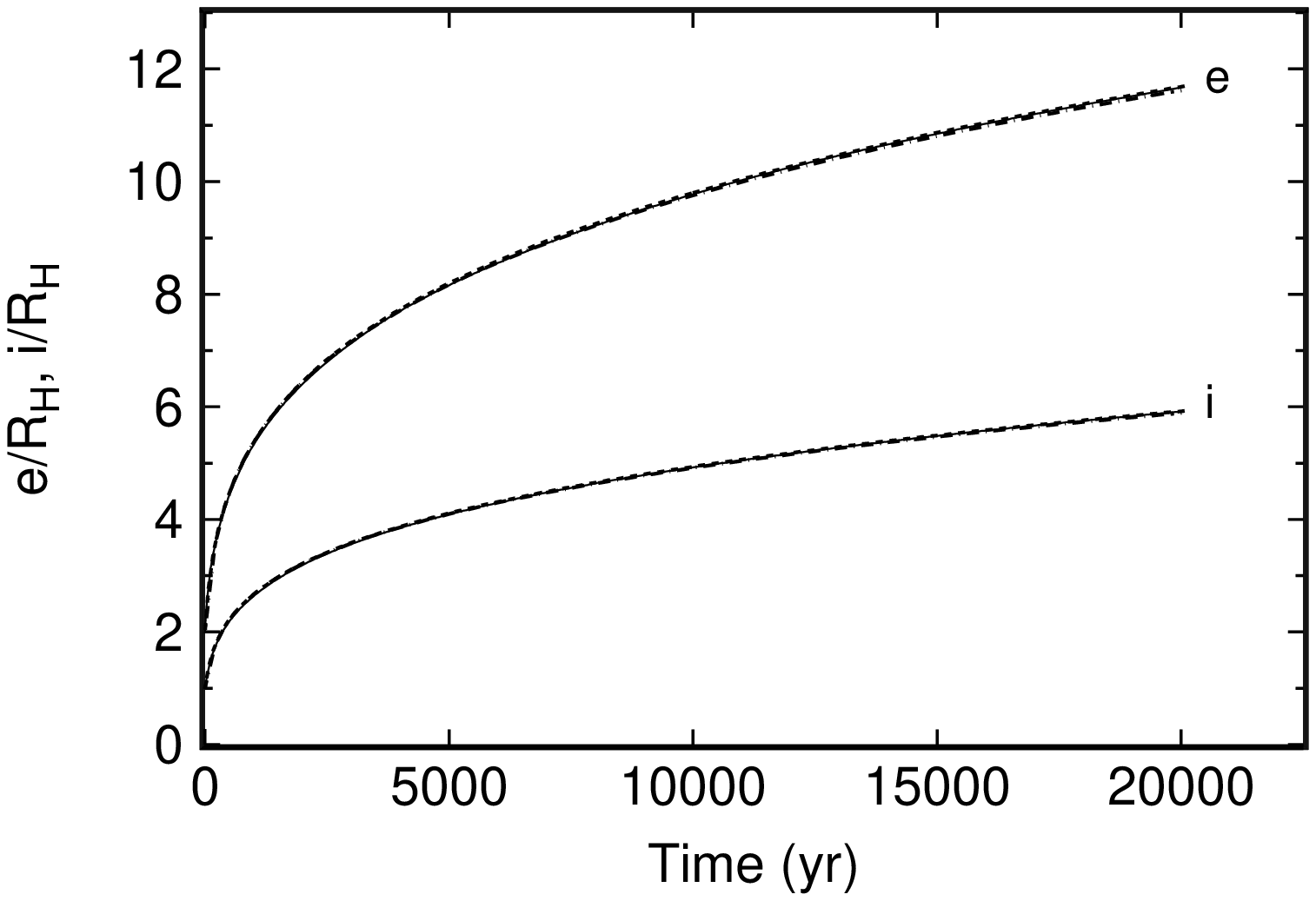}

\figcaption[Kenyon.fig9.ps]
{Evolution of $e$ and $i$ in units of the Hill radius for 800 planetesimals 
with $m_1 = 10^{24}$ g in grids with 20 (solid curves), 40 (dashed curves), 
and 80 (dot-dashed curves) annuli centered at 1 AU.  The surface density 
for each grid is 10 g cm$^{-2}$ at $t = 0$. The evolution of $e$ and $i$ is
independent of the grid spacing.}

\hskip 10ex
\epsffile{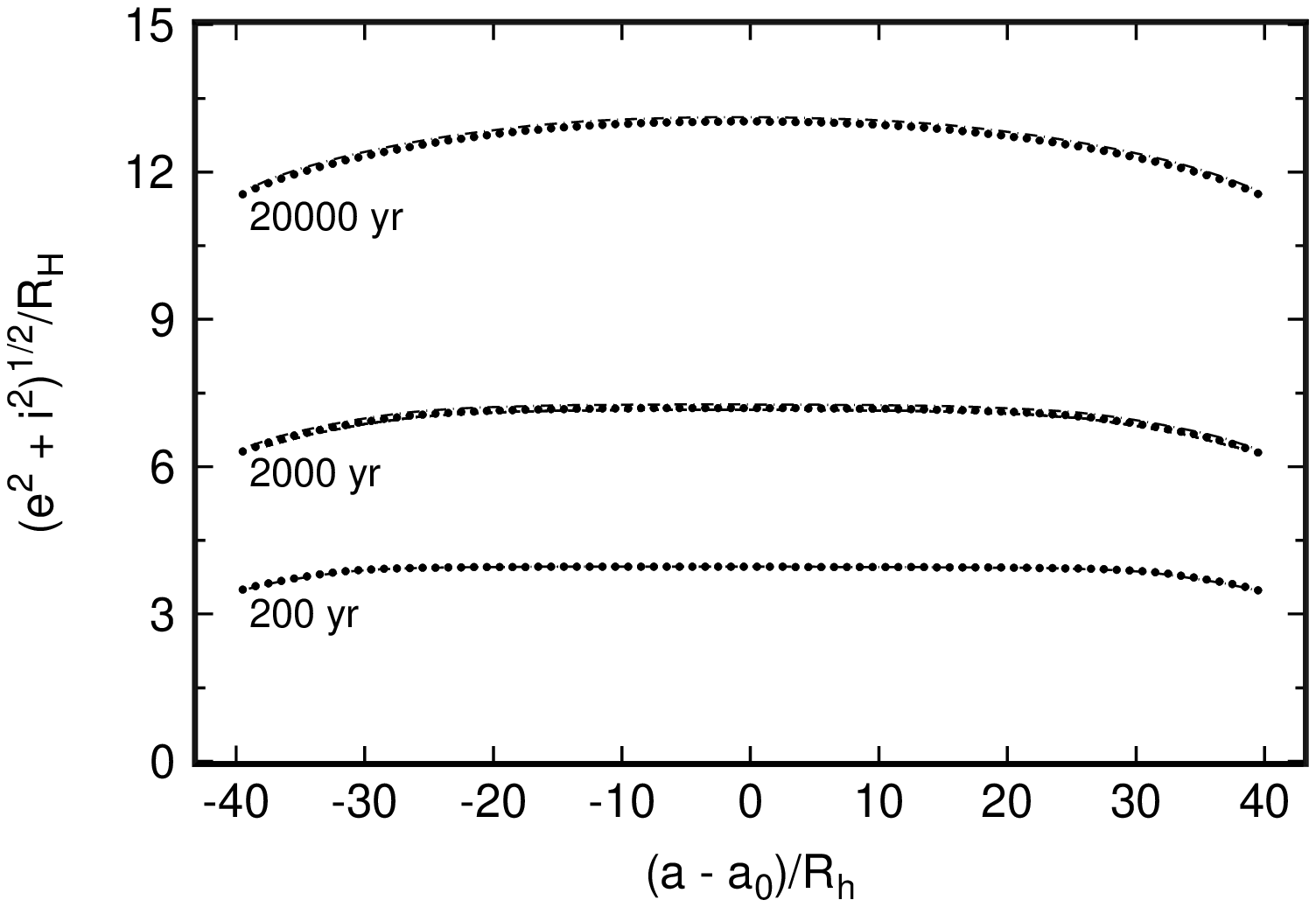}

\figcaption[Kenyon.fig10.ps]
{Evolution of the radial profile of $e^2 + i^2$ for the 800 planetesimal 
calculations in Figure 9. The filled symbols indicate $e^2 + i^2$ for
a grid with 80 annuli. The curves plot results for a grid with 20 annuli
(dashed curves) and for a grid with 40 annuli (dot-dashed curves). 
The evolution time is listed below each profile.}

\hskip 10ex
\epsffile{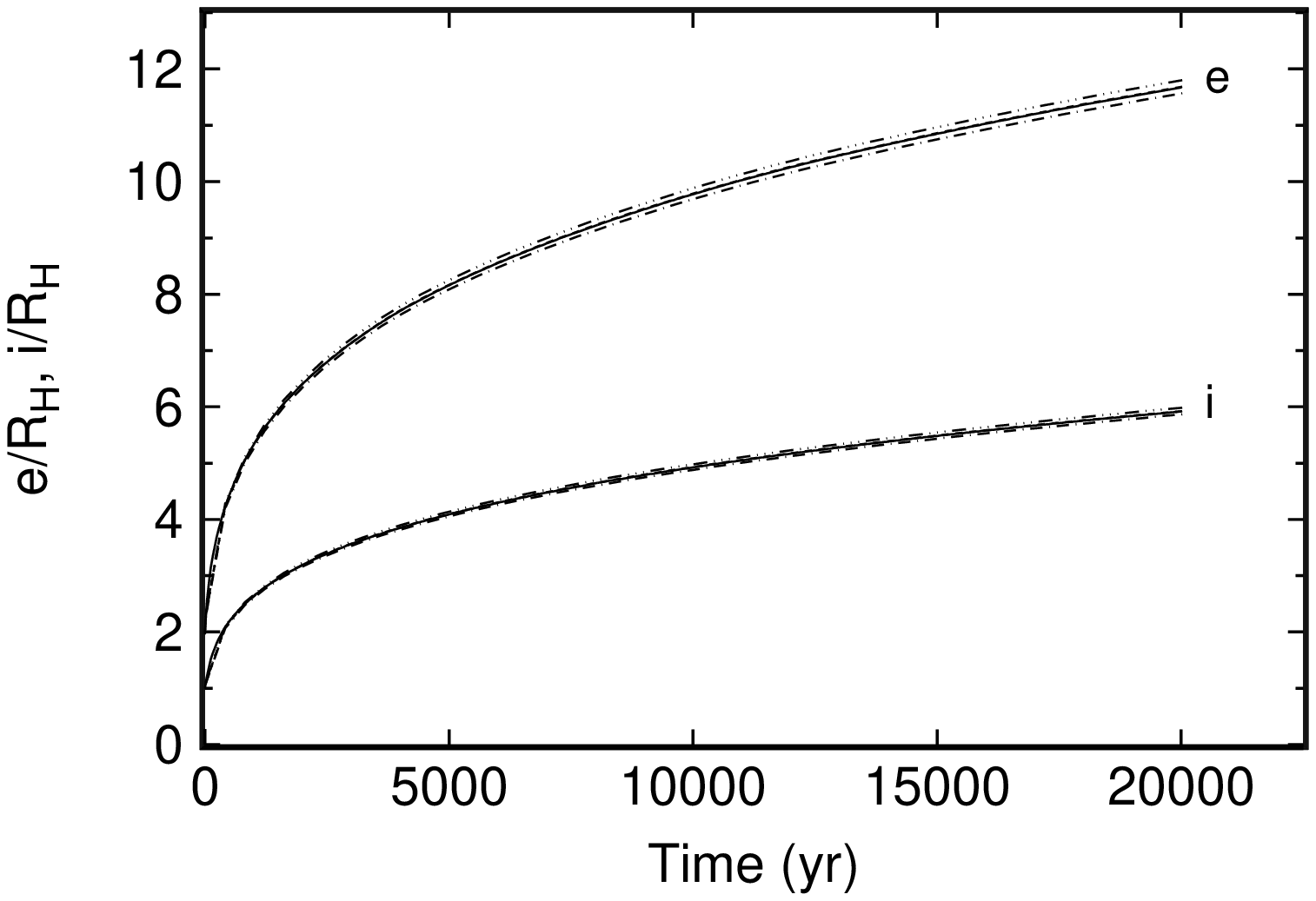}

\figcaption[Kenyon.fig11.ps]
{Scaled evolution of $e$ and $i$ in units of the Hill radius for 
planetesimals in a grid of 40 annuli centered at 1 AU. The four
cases considered are
800 planetesimals with $m_1 = 10^{24}$ g (solid curves),
8000 planetesimals with $m_1 = 10^{24}$ g (dashed curves),
400 planetesimals with $m_1 = 10^{26}$ g (dot-dashed curves), and
$4 \times 10^5$ planetesimals with $m_1 = 10^{23}$ g (dot-dot-dashed curves).
The evolution time has been scaled with the surface density and
planetesimal mass from equations (A14--A15).}

\hskip 10ex
\epsffile{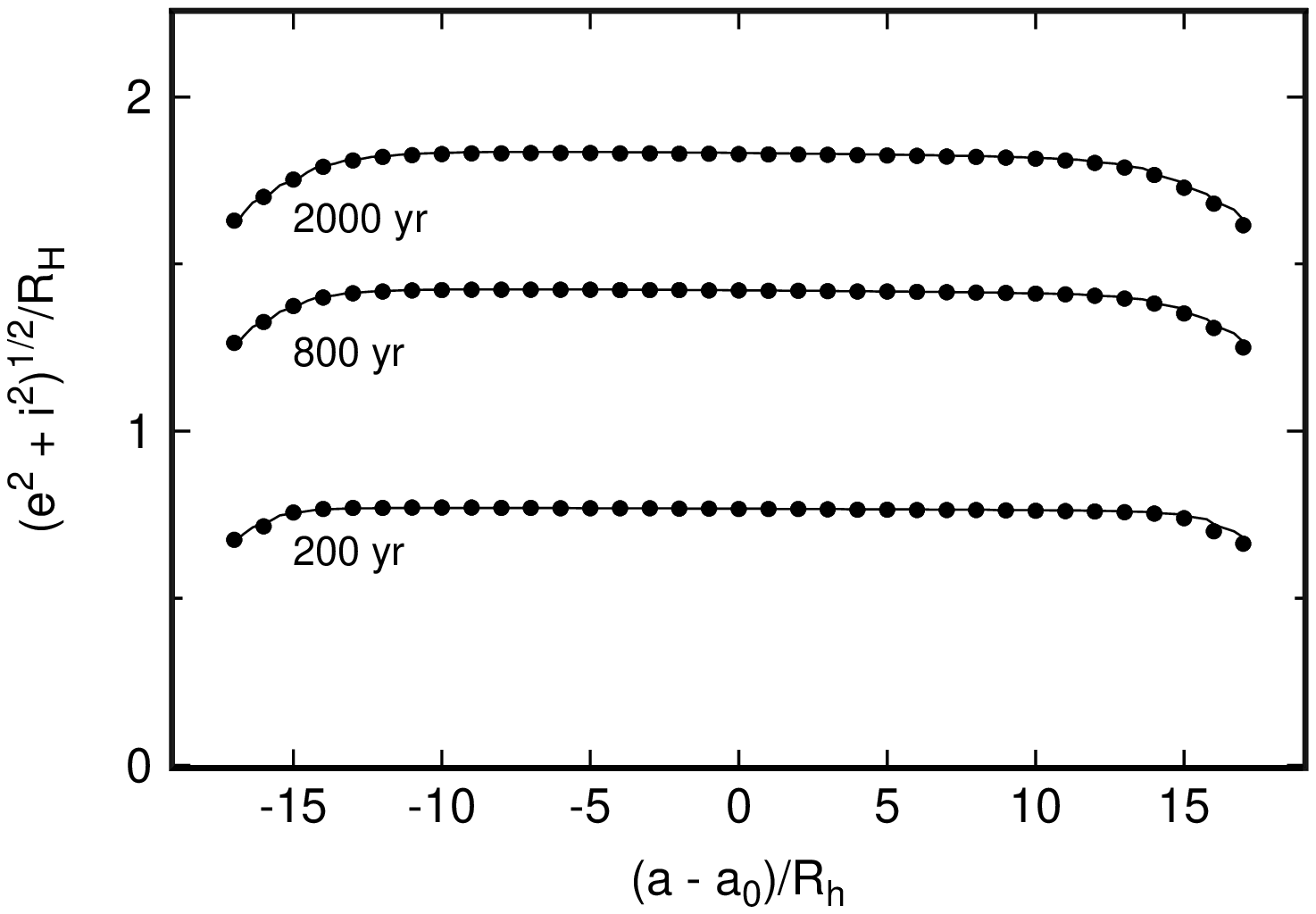}

\figcaption[Kenyon.fig12.ps]
{Evolution of the radial profile of $e^2 + i^2$ for 805 planetesimals 
in a grid centered at 1 AU.  Filled circles indicate results for a grid
of 35 annuli; the solid curve shows results for a grid of 70 annuli.
Both grids extend to $\pm$17.5 $R_H$ from the midpoint as described in
the main text.  The evolution time is listed below each profile.}

\hskip 10ex
\epsffile{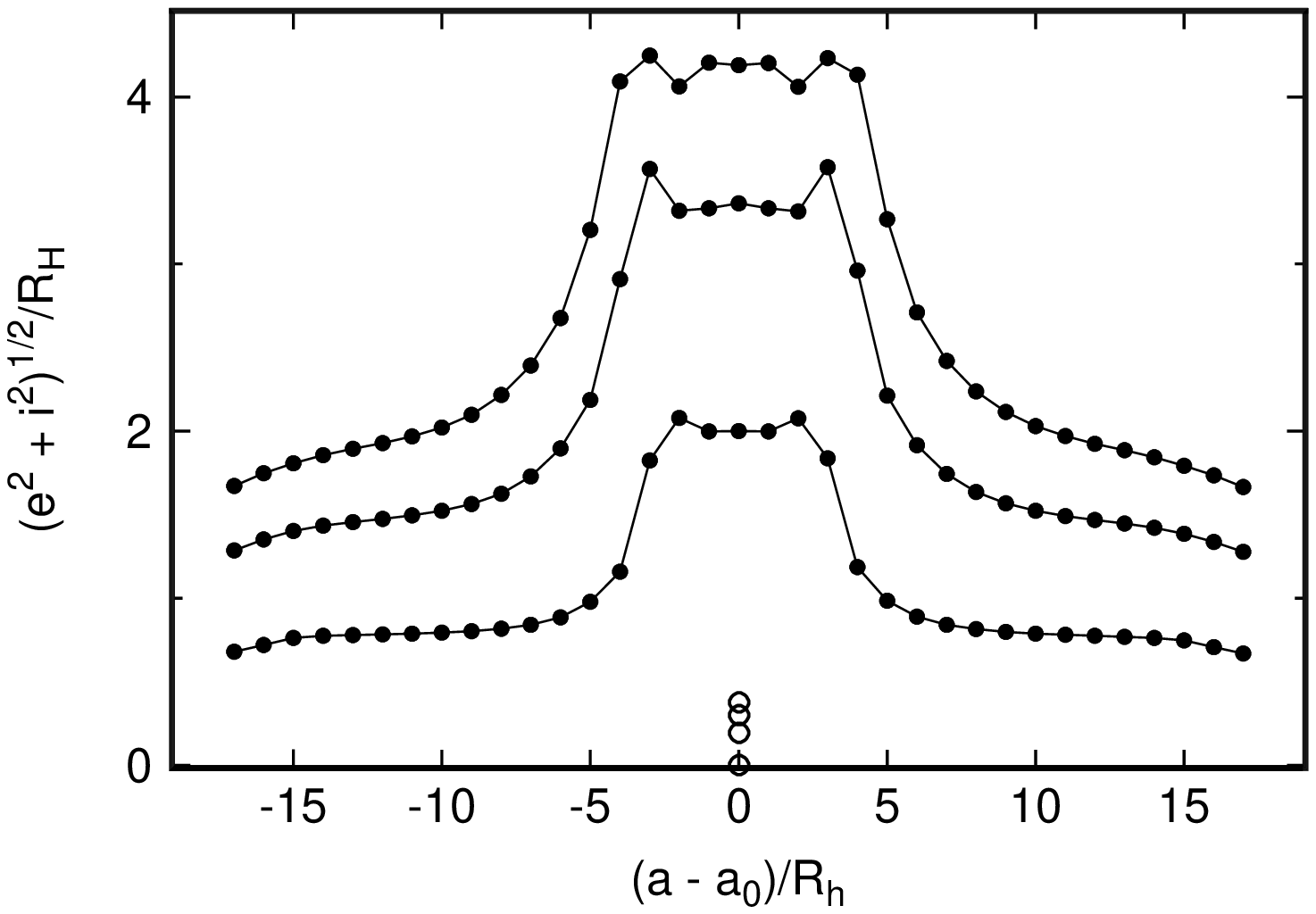}

\figcaption[Kenyon.fig13.ps]
{As in Figure 12, but with an additional planetesimal of mass
$m_2 = 2 \times 10^{26}$ g at $a = a_0$. Open circles indicate
the evolution of the massive planetesimal for $t = $ 0 and other
times as in Figure 12.}

\hskip 10ex
\epsffile{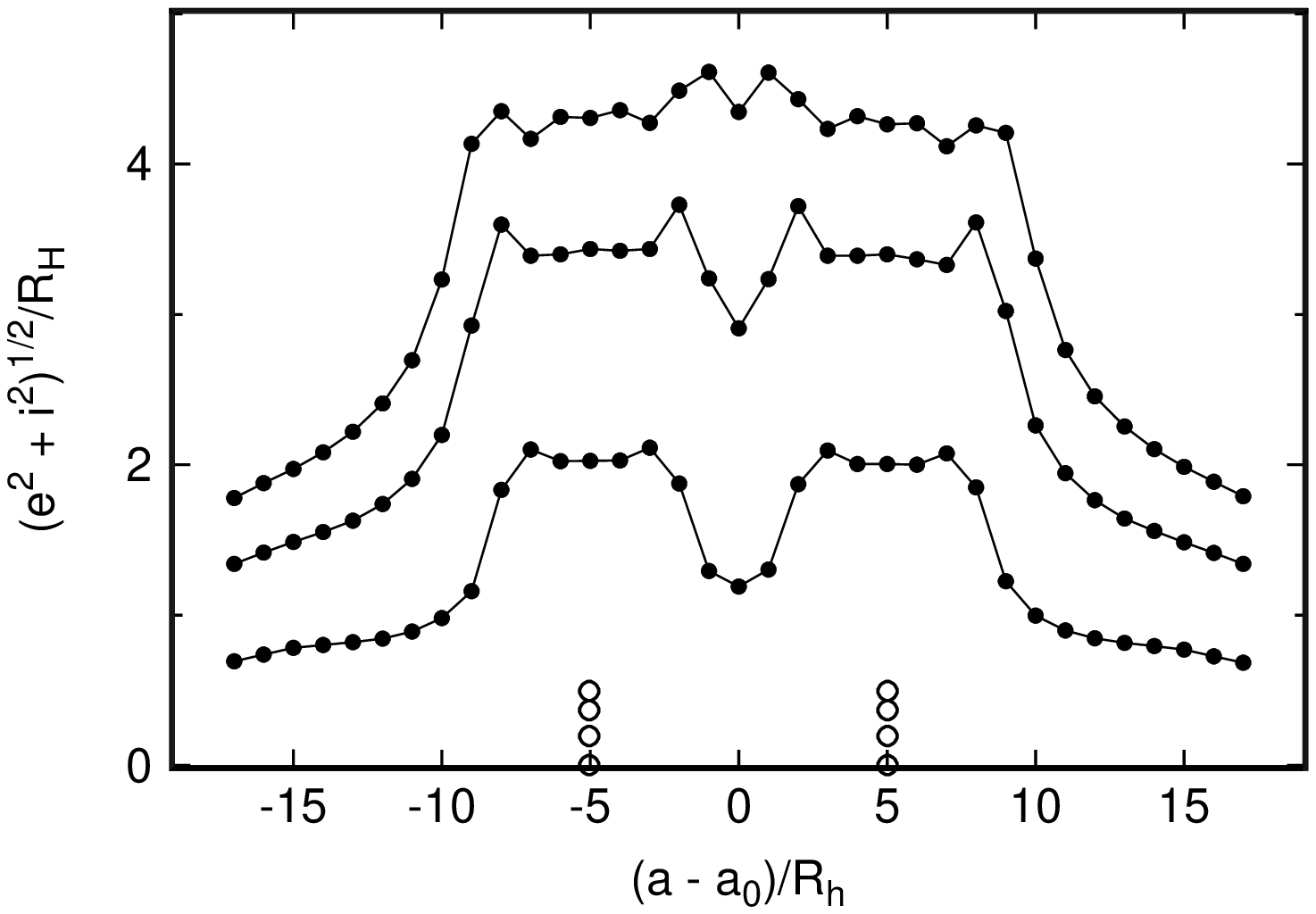}

\figcaption[Kenyon.fig14.ps]
{As in Figure 13, but with two planetesimals of mass $m_2 = $
$2 \times 10^{26}$ g at $a = \pm 5 a_0$. Open circles indicate
the evolution of the massive planetesimals for $t = $ 0 and other
times as in Figure 12.}

\hskip 10ex
\epsffile{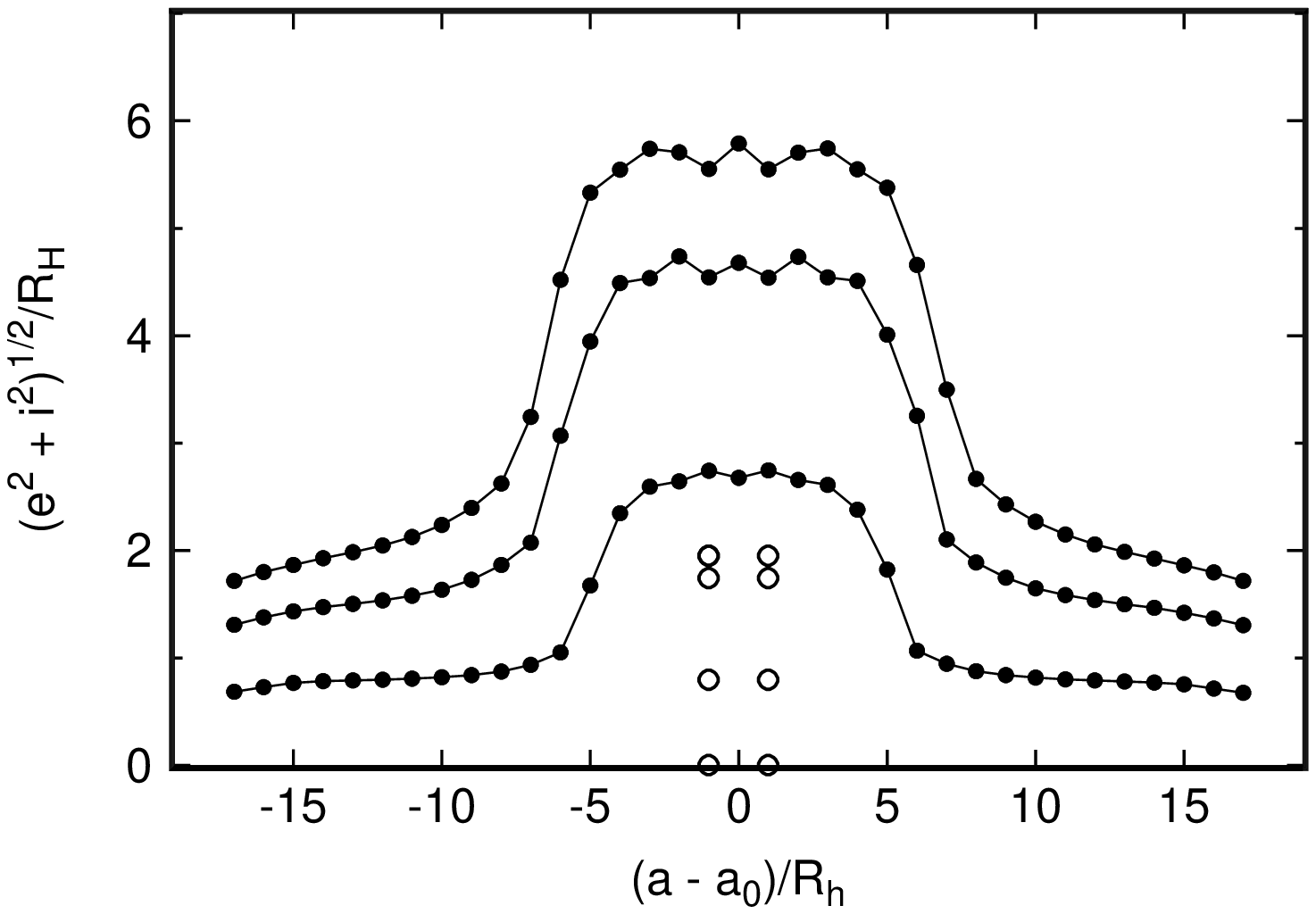}

\figcaption[Kenyon.fig15.ps]
{As in Figure 14, but with two planetesimals $a = \pm 2 a_0$.}

\hskip 10ex
\epsffile{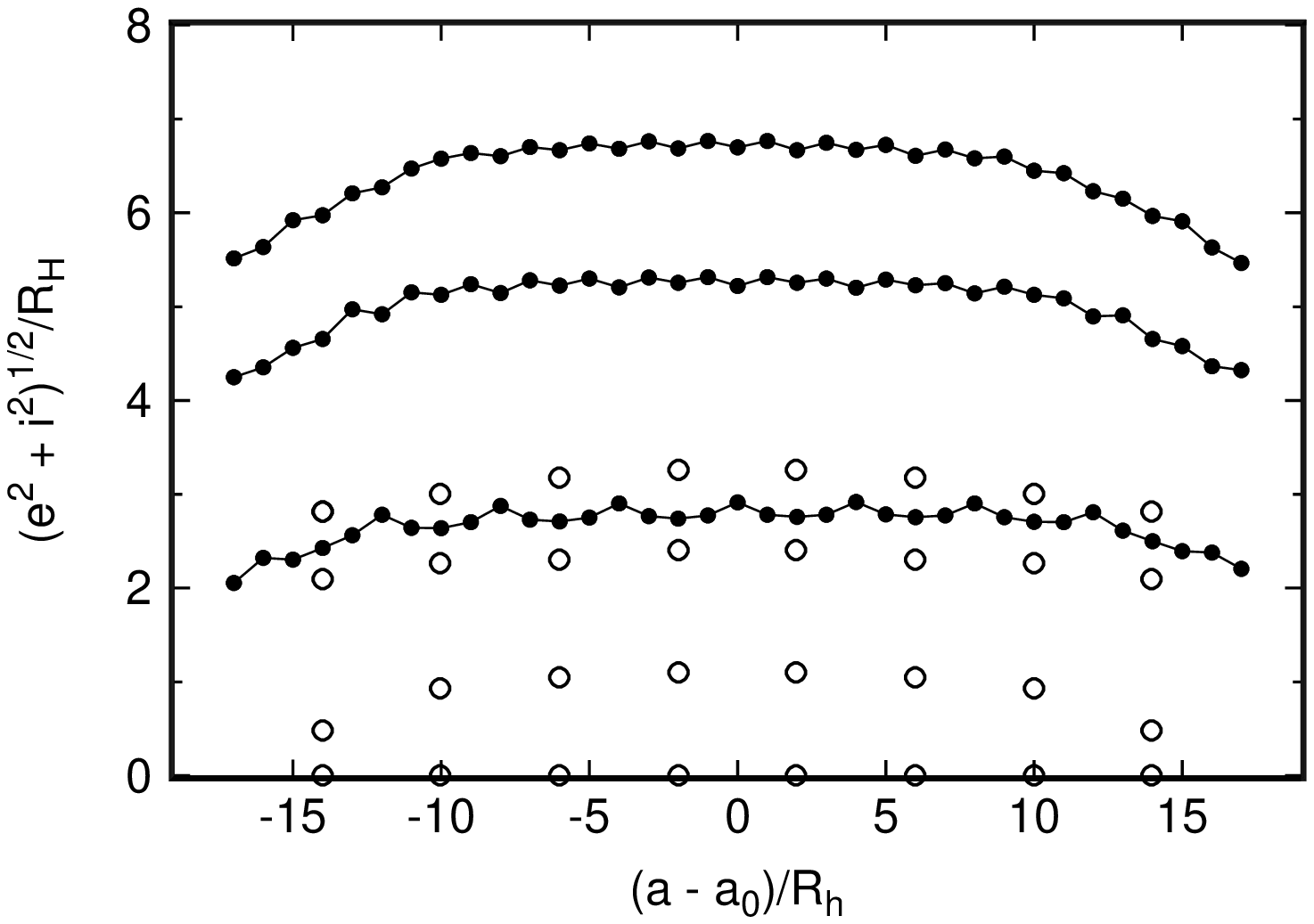}

\figcaption[Kenyon.fig16.ps]
{As in Figure 13, but with eight planetesimals of mass $m_2 = $
$2 \times 10^{26}$ g at $a = \pm 2 a_0$, $\pm 6 a_0$, $\pm 10 a_0$,
and $\pm 14 a_0$.}

\end{document}